\documentclass[english,usenatbib,usegraphicx,usedcolumn,useAMS]{mn2e}
\usepackage{amsmath}
\usepackage{graphicx}
\usepackage{amssymb}
\IfFileExists{url.sty}{\usepackage{url}}
                      {\newcommand{\url}{\texttt}}
\usepackage[authoryear]{natbib}

\makeatletter

\newcommand{\noun}[1]{\textsc{#1}}
\providecommand{\tabularnewline}{\\}

\usepackage{times}
\usepackage{aas_macros}
\newcolumntype{f}{D{.}{.}{5.0}}
\usepackage{hyperref}
\usepackage{hypernat}
\def\elll{L}
\usepackage{babel}
\makeatother
\begin{document}
\title[SDSS galaxy clustering]{SDSS galaxy clustering: luminosity and colour dependence and stochasticity}
\date{Accepted 2008 January 10. Received 2007 December 21; in original form 2007 April 19}
\author[M. E. C. Swanson et al.]{Molly E. C. Swanson,$^{1}$\thanks{E-mail: molly@space.mit.edu} Max Tegmark,$^{1}$ Michael Blanton,$^{2}$ and Idit Zehavi$^{3}$\\  $^{1}$Dept.~of Physics and MIT Kavli Institute, Massachusetts Institute of Technology, 77 Massachusetts Ave, Cambridge, MA 02139,USA\\  $^{2}$Center for Cosmology and Particle Physics, Dept.of Physics, New York University, 4 Washington Pl., New York, NY 10003, USA\\  $^{3}$Dept.of Astronomy, Case Western Reserve University, 10900 Euclid Avenue, Cleveland, OH 44106-7215, USA} 
\maketitle

\begin{abstract}
Differences in clustering properties between galaxy subpopulations
complicate the cosmological interpretation of the galaxy power spectrum,
but can also provide insights about the physics underlying galaxy
formation. To study the nature of this relative clustering, we perform
a counts-in-cells analysis of galaxies in the Sloan Digital Sky Survey
(SDSS) in which we measure the relative bias between pairs of galaxy
subsamples of different luminosities and colours. We use a generalized
$\chi^{2}$ test to determine if the relative bias between each pair
of subsamples is consistent with the simplest deterministic linear
bias model, and we also use a maximum likelihood technique to further
understand the nature of the relative bias between each pair. We find
that the simple, deterministic model is a good fit for the luminosity-dependent
bias on scales above $\sim2\, h^{-1}\rmn{Mpc}$, which is good news
for using magnitude-limited surveys for cosmology. However, the colour-dependent
bias shows evidence for stochasticity and/or non-linearity which increases
in strength toward smaller scales, in agreement with previous studies
of stochastic bias. Also, confirming hints seen in earlier work, the
luminosity-dependent bias for red galaxies is significantly different
from that of blue galaxies: both luminous and dim red galaxies have
higher bias than moderately bright red galaxies, whereas the biasing
of blue galaxies is not strongly luminosity-dependent. These results
can be used to constrain galaxy formation models and also to quantify
how the colour and luminosity selection of a galaxy survey can impact
measurements of the cosmological matter power spectrum. 
\end{abstract}
\begin{keywords} galaxies: statistics -- galaxies: distances and
redshifts -- methods: statistical -- surveys -- large-scale structure
of Universe \end{keywords} %\pacs{98.62.Ve, 98.62.Qz, 98.62.Ai, 98.80.Es}

\section{Introduction}

\label{Introduction} % WHY THIS IS NTERESTING:
In order to use galaxy surveys to study the large-scale distribution
of matter, the relation between the galaxies and the underlying matter
-- known as the \emph{galaxy bias} -- must be understood. Developing
a detailed understanding of this bias is important for two reasons:
bias is a key systematic uncertainty in the inference of cosmological
parameters from galaxy surveys, and it also has implications for galaxy
formation theory.

% CITE EARLIER BIAS MEASUREMENTS:
Since it is difficult to measure the dark matter distribution directly,
we can gain insight by studying \emph{relative bias}, i.e., the relation
between the spatial distributions of different galaxy subpopulations.
There is a rich body of literature on this subject tracing back many
decades (see, e.g., \citealt{1931ApJ....74...43H,1976ApJ...208...13D,1988ApJ...331L..59H,1988ApJ...333L..45W,1994ApJ...431..569P,1995ApJ...442..457L}),
and been studied extensively in recent years as well, both theoretically
\citep{2001MNRAS.325.1359S,2003MNRAS.340..771V,2005MNRAS.363..337C,2005astro.ph.11773S,2007ApJ...659..877T}
and observationally. Such studies have established that biasing depends
on the type of galaxy under consideration -- for example, early-type,
red galaxies are more clustered than late-type, blue galaxies \citep{1997ApJ...489...37G,2002MNRAS.332..827N,2003MNRAS.344..847M,2005MNRAS.356..456C,2006MNRAS.368...21L,2007MNRAS.379.1562C},
and luminous galaxies are more clustered than dim galaxies \citep{1998AJ....115..869W,2001MNRAS.328...64N,2004ApJ...606..702T,2005ApJ...630....1Z,2005PhRvD..71d3511S,2006MNRAS.369...68S}.
Since different types of galaxies do not exactly trace each other,
it is thus impossible for them all to be exact tracers of the underlying
matter distribution.

More quantitatively, the luminosity dependence of bias has been measured
in the 2 Degree Field Galaxy Redshift Survey (2dFGRS; \citealt{2001MNRAS.328.1039C})
\citep{2001MNRAS.328...64N,2002MNRAS.332..827N} and in the Sloan
Digital Sky Survey (SDSS; \citealt{2000AJ....120.1579Y,2002AJ....123..485S})
\citep{2004ApJ...606..702T,2005ApJ...630....1Z,2006MNRAS.368...21L}
as well as other surveys, and it is generally found that luminous
galaxies are more strongly biased, with the difference becoming more
pronounced above $L_{*}$, the characteristic luminosity of a galaxy
in the Schechter luminosity function \citep{1976ApJ...203..297S}.

% WHY OUR APPROACH IS AN IMPROVEMENT GIVING SMALLER ERROR BARS:
These most recent studies measured the bias from ratios of correlation
functions or power spectra. The variances of clustering estimators
like correlation functions and power spectra are well-known to be
the sum of two physically separate contributions: Poisson shot noise
(due to the sampling of the underlying continuous density field with
a finite number of galaxies) and sample variance (due to the fact
that only a finite spatial volume is probed). On the large scales
most relevant to cosmological parameter studies, sample variance dominated
the aforementioned 2dFGRS and SDSS measurements, and therefore dominated
the error bars on the inferred bias.

This sample variance is easy to understand: if the power spectrum
of distant luminous galaxies is measured to be different than that
of nearby dim galaxies, then part of this measured bias could be due
to the nearby region happening to be more/less clumpy than the distant
one. In this paper, we will eliminate this annoying sample variance
by comparing how different galaxies cluster in the \textit{same} region
of space, extending the counts-in-cells work of \citet{1999ApJ...518L..69T},
\citet{2000ApJ...544...63B}, and \citet{2005MNRAS.356..247W} and
the correlation function work of \citet{2001MNRAS.328...64N}, \citet{2002MNRAS.332..827N},
\citet{2005ApJ...630....1Z}, and \citet{2006MNRAS.368...21L}. Here
we use the counts-in-cells technique: we divide the survey volume
into roughly cubical cells and compare the number of galaxies of each
type within each cell. This yields a local, point-by-point measure
of the relative bias rather than a global one as in the correlation
function method. In other words, by comparing two galaxy density fields
directly in real space, including the phase information that correlation
function and power spectrum estimators discard, we are able to provide
substantially sharper bias constraints.

%In this paper we
%tackle the question of the luminosity dependence using counts-in-cells
%rather than calculating the correlation function or the power spectrum,
%and by separating galaxies by colour as well we find that the observed
%behaviour depends strongly on the colour selection.
%IT'S STILL TOO EARLY IN THE PAPER TO SAY WHAT WE FIND.

% OUR APPROACH ALLOWS CONSTRAINING r AS WELL:
This local approach also enables us to quantify so-called stochastic
bias \citep{1998ApJ...504..601P,1998ApJ...500L..79T,1999ApJ...520...24D,1999ApJ...525..543M}.
It is well-known that the relation between galaxies and dark matter
or between two different types of galaxies is not necessarily deterministic
-- galaxy formation processes that depend on variables other than
the local matter density give rise to stochastic bias as described
in \citet{1998ApJ...504..601P}, \citet{1998ApJ...500L..79T}, \citet{1999ApJ...520...24D},
and \citet{1999ApJ...525..543M}. Evidence for stochasticity in the
relative bias between early-type and late-type galaxies has been presented
in \citet{2005MNRAS.356..247W}, \citet{2005MNRAS.356..456C}, \citet{1999ApJ...518L..69T},
and \citet{2000ApJ...544...63B}. Additionally, \citet{2007A&A...461..861S}
finds evidence for stochastic bias between galaxies and dark matter
via weak lensing. The time evolution of such stochastic bias has been
modelled in \citet{1998ApJ...500L..79T} and was recently updated
in \citet{2005A&A...430..827S}. Stochasticity is even predicted in
the relative bias between virialized clumps of dark matter (haloes)
and the linearly-evolved dark matter distribution \citep{2002MNRAS.333..730C,2004MNRAS.355..129S}.
Here we aim to test whether stochasticity is necessary for modelling
the luminosity-dependent or the colour-dependent relative bias.

% WHAT WE'LL DO AND WHY IT'S TIMELY:
In this paper, we study the relative bias as a function of scale using
a simple stochastic biasing model by comparing pairs of SDSS galaxy
subsamples in cells of varying size. Such a study is timely for two
reasons. First of all, the galaxy power spectrum has recently been
measured to high precision on large scales with the goal of constraining
cosmology \citep{2004ApJ...606..702T,2006PhRvD..74l3507T,2007MNRAS.374.1527B,2007MNRAS.378..852P}.
As techniques continue to improve and survey volumes continue to grow,
it is necessary to reduce systematic uncertainties in order to keep
pace with shrinking statistical uncertainties. A detailed understanding
of complications due to the dependence of galaxy bias on scale, luminosity,
and colour will be essential for making precise cosmological inferences
with the next generation of galaxy redshift surveys \citep{2004MNRAS.347..645P,2005ApJ...625..613A,2007ApJ...657..645P,2007ApJ...659....1Z,2007MNRAS.379.1195M,2007PhRvD..75h3510K}.

Secondly, a great deal of theoretical progress on models of galaxy
formation has been made in recent years, and 2dFGRS and SDSS contain
a large enough sample of galaxies that we can now begin to place robust
and detailed observational constraints on these models. The framework
known as the halo model \citep{2000MNRAS.318..203S} (see \citealt{2002PhR...372....1C}
for a comprehensive review) provides the tools needed to make comparisons
between theory and observations. The halo model assumes that all galaxies
form in dark matter haloes, so the galaxy distribution can be modelled
by first determining the halo distribution -- either analytically
\citep{1998MNRAS.297..692C,2006PhRvD..74j3512M,2007PhRvD..75f3512S}
or using $N$-body simulations \citep{2003MNRAS.341.1311S,2004ApJ...609...35K,2006astro.ph..7463K}
-- and then populating the haloes with galaxies. This second step
can be done using semi-analytical galaxy formation models \citep{2001MNRAS.320..289S,2003ApJ...593....1B,2006MNRAS.365...11C,2006RPPh...69.3101B}
or with a statistical approach using a model for the halo occupation
distribution (HOD) \citep{2000MNRAS.318.1144P,2002ApJ...575..587B,2005PhRvD..71f3001S}
or conditional luminosity function (CLF) \citep{2003MNRAS.339.1057Y,2003MNRAS.340..771V}
which prescribes how galaxies populate haloes.

Although there are some concerns that the halo model does not capture
all of the relevant physics \citep{2006ApJ...638L..55Y,2006astro.ph..1090C,2007MNRAS.377L...5G},
it has been applied successfully in a number of different contexts
\citep{2003MNRAS.339..410S,2005MNRAS.361..415C,2006ApJ...647..737T,2006MNRAS.369...68S}.
The correlation between a galaxy's environment (i.e., the local density
of surrounding galaxies) and its colour and luminosity (\citealt{2004ApJ...601L..29H};
Blanton et al. 2005a)\nocite{2005ApJ...629..143B} has been interpreted
in the context of the halo model \citep{2005ApJ...629..625B,2006ApJ...645..977B,2006MNRAS.372.1749A},
and \citet{2003MNRAS.340..771V} and \citet{2005MNRAS.363..337C}
make predictions for the bias as a function of galaxy type and luminosity
using the CLF formalism. Additionally, \citet{2005ApJ...630....1Z},
\citet{2003MNRAS.346..186M}, and \citet{2005MNRAS.364.1327A} use
correlation function methods to study the luminosity and colour dependence
of galaxy clustering, and interpret the results using the Halo Occupation
Distribution (HOD) framework. The analysis presented here is complementary
to this body of work in that the counts-in-cells method is sensitive
to larger scales, uses a different set of assumptions, and compares
the two density fields directly in each cell rather than comparing
ratios of their second moments. The halo model provides a natural
framework in which to interpret the luminosity and colour dependence
of galaxy biasing statistics we measure here.

The rest of this paper is organized as follows: Section~\ref{SDSS-Galaxy-Data}
describes our galaxy data, and Sections~\ref{sub:Overlapping-Volume-Limited-Samples}~and~\ref{sub:Counts-in-Cells-Methodology}
describe the construction of our galaxy samples and the partition
of the survey volume into cells. In Section~\ref{sub:Relative-Bias-Framework}
we outline our relative bias framework, and in Sections~\ref{sub:The-Null-buster-Test}~and~\ref{sub:Maximum-Likelihood-Method}
we describe our two main analysis methods. We present our results
in Section~\ref{Results} and conclude with a qualitative interpretation
of our results in the halo model context in Section~\ref{Conclusions}.

\section{SDSS Galaxy Data}

\label{SDSS-Galaxy-Data} The SDSS \citep{2000AJ....120.1579Y,2002AJ....123..485S}
uses a mosaic CCD camera \citep{1998AJ....116.3040G} on a dedicated
telescope \citep{2006AJ....131.2332G} to image the sky in five photometric
bandpasses denoted $u$, $g$, $r$, $i$ and $z$ \citep{1996AJ....111.1748F}.
After astrometric calibration \citep{2003AJ....125.1559P}, photometric
data reduction \citep{2001ASPC..238..269L}, and photometric calibration
\citep{2001AJ....122.2129H,2002AJ....123.2121S,2004AN....325..583I,2006AN....327..821T},
galaxies are selected for spectroscopic observations. To a good approximation,
the main galaxy sample consists of all galaxies with $r$-band apparent
Petrosian magnitude $r<17.77$ after correction for reddening as per
\citet{1998ApJ...500..525S}; there are about 90 such galaxies per
square degree, with a median redshift of 0.1 and a tail out to $z\sim0.25$.
Galaxy spectra are also measured for the Luminous Red Galaxy sample
\citep{2001AJ....122.2267E}, which is not used in this paper. These
targets are assigned to spectroscopic plates of diameter $2.98^{\circ}$
by an adaptive tiling algorithm (Blanton et al. 2003b)\nocite{2003AJ....125.2276B}
and observed with a pair of CCD spectrographs \citep{2004SPIE.5492.1411U},
after which the spectroscopic data reduction and redshift determination
are performed by automated pipelines. The rms galaxy redshift errors
are of order $30\,\rmn{km\, s^{-1}}$ for main galaxies, hence negligible
for the purposes of the present paper.

Our analysis is based on 380,614 main galaxies (the `safe0' cut) from
the 444,189 galaxies in the 5th SDSS data release (`DR5') \citep{2007ApJS..172..634A},
processed via the SDSS data repository at New York University (Blanton
et al. 2005b)\nocite{2005AJ....129.2562B}. The details of how these
samples were processed and modelled are given in Appendix A of \citet{2004ApJ...606..702T}
and in \citet{2005ApJ...633..560E}. The bottom line is that each
sample is completely specified by three entities:

\begin{enumerate}
\item The galaxy positions (RA, Dec, and comoving redshift space distance
$r$ for each galaxy) ;
\item The radial selection function $\bar{n}\left(r\right)$, which gives
the expected (not observed) number density of galaxies as a function
of distance ;
\item The angular selection function $\bar{n}\left(\hat{\bmath{r}}\right)$,
which gives the completeness as a function of direction in the sky,
specified in a set of spherical polygons \citep{2004MNRAS.349..115H}. 
\end{enumerate}
The three-dimensional selection functions of our samples are separable,
i.e., simply the product $\bar{n}\left(\bmath{r}\right)=\bar{n}\left(\hat{\bmath{r}}\right)\bar{n}\left(r\right)$
of an angular and a radial part; here $r\equiv|\bmath{r}|$ and $\hat{\bmath{r}}\equiv\bmath{r}/r$
are the radial comoving distance and the unit vector corresponding
to the position $\bmath{r}$. The volume-limited samples used in this
paper are constructed so that their radial selection function $\bar{n}\left(r\right)$
is constant over a range of $r$ and zero elsewhere. The effective
sky area covered is $\Omega\equiv\int\bar{n}\left(\hat{\bmath{r}}\right)d\Omega\approx5183$
square degrees, and the typical completeness $\bar{n}\left(\hat{\bmath{r}}\right)$
exceeds 90 per cent. The conversion from redshift $z$ to comoving
distance was made for a flat $\Lambda\rmn{CDM}$ cosmological model
with $\Omega_{m}=0.25$. %{[}effect on results?] 
% MT: I suggest not going into this unless the referee requests it. 
% Although it has a (quite small) effect on the recovered P(k) due to projection effects,
% the 3D maps of all galaxy subpopulations of course get stretched the same way, so it shouldn't have any
% impact on our measurements of b and r; changing the fiducial Omega_m would be equivalent to simply shifting the
% radial pixel boundaries.
Additionally, we make a first-order correction for redshift space
distortions by applying the finger-of-god compression algorithm described
in \citet{2004ApJ...606..702T} with a threshold density of $\delta_{c}=200$.
% MT: I THINK THE FOG COMPRESSION IS MORE LIKELY TO AFFECT THE RESULTS THAN VARIATION OF THE FIGUCAL Omega_m.

\section{Analysis Methods}

\label{Analysis-Methods}

\subsection{\label{sub:Overlapping-Volume-Limited-Samples}Overlapping Volume-Limited
Samples}

The basic technique used in this paper is pairwise comparison of the
three-dimensional density fields of galaxy samples with different
colours and luminosities. As in \citet{2005ApJ...630....1Z}, we focus
on these two properties (as opposed to morphological type, spectral
type, or surface brightness) for two reasons: they are straightforward
to measure from SDSS data, and recent work (Blanton et al. 2005a)
has found that luminosity and colour is the pair of properties that
is most predictive of the local overdensity. Since colour and spectral
type are strongly correlated, our study of the colour dependence of
bias probes similar physics as studies using spectral type \citep{1999ApJ...518L..69T,2000ApJ...544...63B,2002MNRAS.332..827N,2005MNRAS.356..247W,2005MNRAS.356..456C}.

Our base sample of SDSS galaxies (`safe0') has an $r$-band apparent
magnitude range of $14.5<r<17.5$. Following the method used in \citet{2004ApJ...606..702T},
we created a series of volume-limited samples containing galaxies
in different luminosity ranges. These samples are defined by selecting
a range of absolute magnitude $M_{\rmn{luminous}}<M_{^{0.1}r}<M_{\rmn{dim}}$
and defining a redshift range such that the near limit has $M_{^{0.1}r}=M_{\rmn{luminous}}$,
$r=14.5$ and the far limit has $M_{^{0.1}r}=M_{\rmn{dim}}$, $r=17.5$.
Thus by discarding all galaxies outside the redshift range, we are
left with a sample with a uniform radial selection function $\bar{n}\left(r\right)$
that contains all of the galaxies in the given absolute magnitude
range in the volume defined by the redshift limits. Here $M_{^{0.1}r}$
is defined as the absolute magnitude in the $r$-band shifted to a
redshift of $z=0.1$ (Blanton et al. 2003a)\nocite{2003AJ....125.2348B}.
\begin{figure}
\includegraphics[width=1\columnwidth]{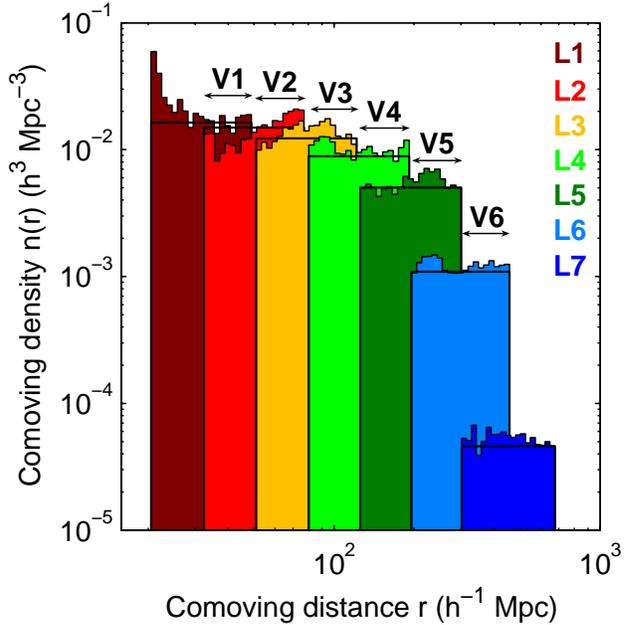}

\caption{\label{fig:vlim_hist}Histogram of the comoving number density (after
finger-of-god compression) of the volume-limited samples L1-L7. The
cuts used to define these samples are shown in Table~\ref{tab:cuts}.
Note that the radial selection function $\bar{n}\left(r\right)$ is
uniform over the allowed range for each sample. Arrows indicate volumes
V1-V6 where neighbouring volume-limited samples overlap.}
\end{figure}

Our volume-limited samples are labelled L1 through L7, with L1 being
the dimmest and L7 being the most luminous. Figure~\ref{fig:vlim_hist}
shows a histogram of the comoving galaxy density $n\left(r\right)$
for L1-L7. The cuts used to make these samples are shown in Table~\ref{tab:cuts}.%
\begin{table*}
\begin{minipage}{126mm}

\caption{\label{tab:cuts}Summary of cuts used to create luminosity-binned
volume-limited samples.}

\begin{tabular}{cccc}
\hline 
Luminosity-binned&
&
&
Comoving number\tabularnewline
volume-limited samples&
Absolute magnitude&
Redshift&
density $\bar{n}$ $\left(h^{3}{\rmn{Mpc}}^{-3}\right)$\tabularnewline
\hline 
L1&
$-17<M_{^{0.1}r}<-16$&
$0.007<z<0.016$&
$\left(1.63\pm0.05\right)\times10^{-2}$\tabularnewline
L2&
$-18<M_{^{0.1}r}<-17$&
$0.011<z<0.026$&
$\left(1.50\pm0.03\right)\times10^{-2}$\tabularnewline
L3&
$-19<M_{^{0.1}r}<-18$&
$0.017<z<0.041$&
$\left(1.23\pm0.01\right)\times10^{-2}$\tabularnewline
L4&
$-20<M_{^{0.1}r}<-19$&
$0.027<z<0.064$&
$\left(8.86\pm0.05\right)\times10^{-3}$\tabularnewline
L5&
$-21<M_{^{0.1}r}<-20$&
$0.042<z<0.100$&
$\left(5.02\pm0.02\right)\times10^{-3}$\tabularnewline
L6&
$-22<M_{^{0.1}r}<-21$&
$0.065<z<0.152$&
$\left(1.089\pm0.005\right)\times10^{-3}$\tabularnewline
L7&
$-23<M_{0.1_{r}}<-22$&
$0.101<z<0.226$&
$\left(4.60\pm0.06\right)\times10^{-5}$\tabularnewline
\hline
\end{tabular}

\end{minipage} 
\end{table*}

Each sample overlaps spatially only with the samples in neighbouring
luminosity bins -- since the apparent magnitude range spans three
magnitudes and the absolute magnitude ranges for each bin span one
magnitude, the far redshift limit of a given luminosity bin is approximately
equal to the near redshift limit of the bin two notches more luminous.
(It is not precisely equal due to evolution and K-corrections.)

The regions where neighbouring volume-limited samples overlap provide
a clean way to select data for studying the luminosity-dependent bias.
By using only the galaxies in the overlapping region from each of
the two neighbouring luminosity bins, we obtain two sets of objects
(one from the dimmer bin and one from more luminous bin) whose selection
is volume-limited and redshift-independent. Furthermore, since they
occupy the same volume, they are correlated with the same underlying
matter distribution, which eliminates uncertainty due to sample variance
and removes potential systematic effects due to sampling different
volume sizes \citep{2005A&A...443...11J}. We label the overlapping
volume regions V1 through V6, where V1 is defined as the overlap between
L1 and L2, and so forth. The redshift ranges for V1-V6 are shown Table~\ref{tab:vols}.%
\begin{table}

\caption{\label{tab:vols}Overlapping volumes in which neighbouring luminosity
bins are compared.}

\begin{tabular}{ccc}
\hline 
Pairwise comparison &
Overlapping&
\tabularnewline
(overlapping) volumes&
bins&
Redshift\tabularnewline
\hline 
V1&
L1 \& L2&
$0.011<z<0.016$\tabularnewline
V2&
L2 \& L3&
$0.017<z<0.026$\tabularnewline
V3&
L3 \& L4&
$0.027<z<0.041$\tabularnewline
V4&
L4 \& L5&
$0.042<z<0.064$\tabularnewline
V5&
L5 \& L6&
$0.065<z<0.100$\tabularnewline
V6&
L6 \& L7&
$0.101<z<0.152$\tabularnewline
\hline
\end{tabular}
\end{table}

\begin{figure}
\includegraphics[width=1\columnwidth]{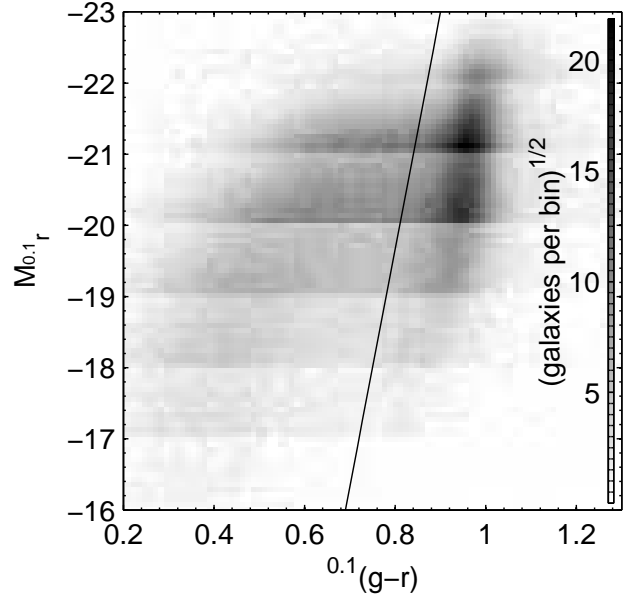}

\caption{\label{fig:colourmag}Colour-magnitude diagram showing the number
density distribution of the galaxies in the volume-limited samples.
The shading scale has a square-root stretch, with darker areas indicating
regions of higher density. The line shows the colour cut of $^{0.1}\left(g-r\right)=0.9-0.03\left(M_{^{0.1}r}+23\right)$.
We refer to galaxies falling to the left of this line as blue and
ones falling to the right of the line as red. }
\end{figure}

To study the colour dependence of the bias, we further divide each
sample into red galaxies and blue galaxies. Figure~\ref{fig:colourmag}
shows the galaxy distribution of our volume-limited samples on a colour-magnitude
diagram. The sharp boundaries between the different horizontal slices
are due to the differences in density and total volume sampled in
each luminosity bin. This diagram illustrates the well-known colour
bimodality, with the redder galaxies falling predominantly in a region
commonly known as the E-S0 ridgeline \citep{2004ApJ...615L.101B,2006MNRAS.370..721M,2006MNRAS.373..469B}.
To separate the E-S0 ridgeline from the rest of the population, we
use the same magnitude-dependent colour cut as in \citet{2005ApJ...630....1Z}:
we define galaxies with $^{0.1}\left(g-r\right)<0.9-0.03\left(M_{^{0.1}r}+23\right)$
to be blue and galaxies on the other side of this line to be red.%

\begin{figure*}
\includegraphics[width=1\textwidth]{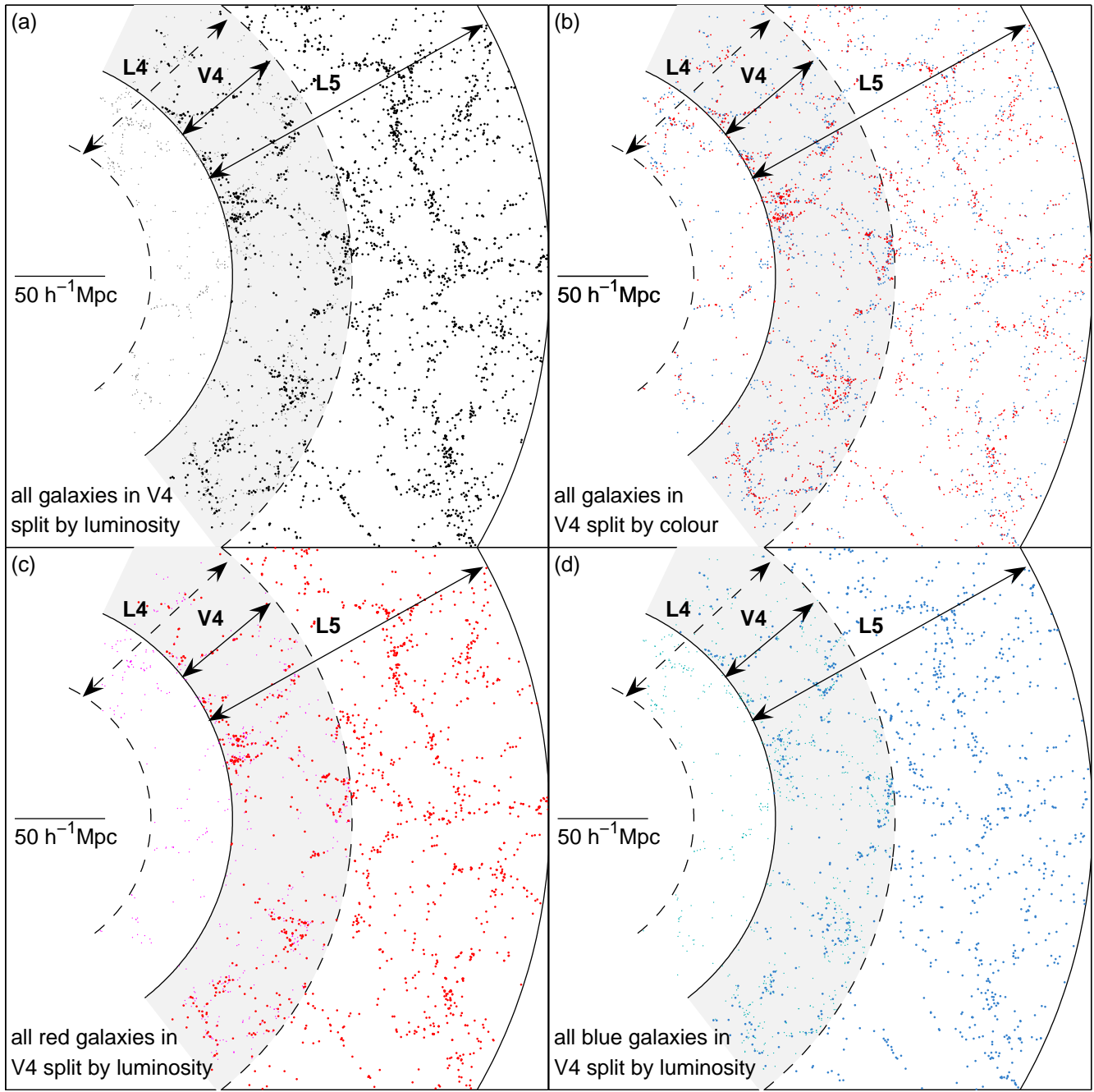}

\caption{\label{fig:4pairwise}Galaxy distributions (after finger-of-god compression)
plotted in comoving spatial co-ordinates for a radial slice of the
volume-limited samples L4 (smaller dots, radial boundaries denoted
by dashed lines) and L5 (larger dots, radial boundaries denoted by
solid lines), which overlap in volume V4. Four different types of
pairwise comparisons are illustrated: (a) luminous galaxies (L5) vs.
dim galaxies (L4), (b) red galaxies vs. blue galaxies (both in V4),
(c) luminous red galaxies (L5) vs. dim red galaxies (L4), and (d)
luminous blue galaxies (L5) vs. dim blue galaxies (L4). The shaded
regions denote the volume in which the two sets of galaxies are compared.
A simple visual inspection shows that the different samples of galaxies
being compared generally appear to cluster in the same physical locations
-- one key question we aim to answer here is if these observed correlations
can be described with a simple linear bias model.}
\end{figure*}

In each volume V1-V6, we make four separate pairwise comparisons:
luminous galaxies vs. dim galaxies, red galaxies vs. blue galaxies,
luminous red galaxies vs. dim red galaxies, and luminous blue galaxies
vs. dim blue galaxies. The luminous vs. dim comparisons measures the
relative bias between galaxies in neighbouring luminosity bins, and
from this we can extract the luminosity dependence of the bias for
all galaxies combined and for red and blue galaxies separately. The
red vs. blue comparison measures the colour-dependent bias. This set
of four different types of pairwise comparisons is illustrated in
Fig.~\ref{fig:4pairwise} for V4, and the number of galaxies in each
sample being compared is shown in Table~\ref{tab:counts}.%
\begin{table}

\caption{\label{tab:counts}Number of galaxies in each sample being compared.}

\begin{tabular}{c|ff||ff}
\hline 
&
\multicolumn{2}{c||}{All split by luminosity}&
\multicolumn{2}{c}{All split by colour}\tabularnewline
&
\multicolumn{1}{c}{Luminous}&
\multicolumn{1}{c||}{Dim}&
\multicolumn{1}{c}{Red}&
\multicolumn{1}{c}{Blue}\tabularnewline
\hline 
V1&
427&
651&
125&
953\tabularnewline
V2&
2102&
2806&
1117&
3791\tabularnewline
V3&
6124&
8273&
5147&
9250\tabularnewline
V4&
12122&
23534&
17144&
18512\tabularnewline
V5&
11202&
53410&
37472&
27140\tabularnewline
V6&
1784&
38920&
27138&
13566\tabularnewline
\hline
\hline 
&
\multicolumn{2}{c||}{Red split by luminosity}&
\multicolumn{2}{c}{Blue split by luminosity}\tabularnewline
&
\multicolumn{1}{c}{Red luminous}&
\multicolumn{1}{c||}{Red dim}&
\multicolumn{1}{c}{Blue luminous}&
\multicolumn{1}{c}{Blue dim}\tabularnewline
\hline 
V1&
72&
53&
355&
598\tabularnewline
V2&
620&
497&
1482&
2309\tabularnewline
V3&
2797&
2350&
3327&
5923\tabularnewline
V4&
6848&
10296&
5274&
13238\tabularnewline
V5&
7514&
29958&
3688&
23452\tabularnewline
V6&
1451&
25687&
333&
13233\tabularnewline
\hline
\end{tabular}
\end{table}

\subsection{\label{sub:Counts-in-Cells-Methodology}Counts-in-Cells Methodology}

To compare the different pairs of galaxy samples, we perform a counts-in-cells
analysis: we divide each comparison volume into roughly cubical cells
and use the number of galaxies of each type in each cell as the primary
input to our statistical analysis. This method is complementary to
studies based on the correlation function since it involves point-by-point
comparison of the two density fields and thus provides a more direct
test of the local deterministic linear bias hypothesis. We probe scale
dependence by varying the size of the cells.

\begin{figure*}
\includegraphics[width=1\textwidth]{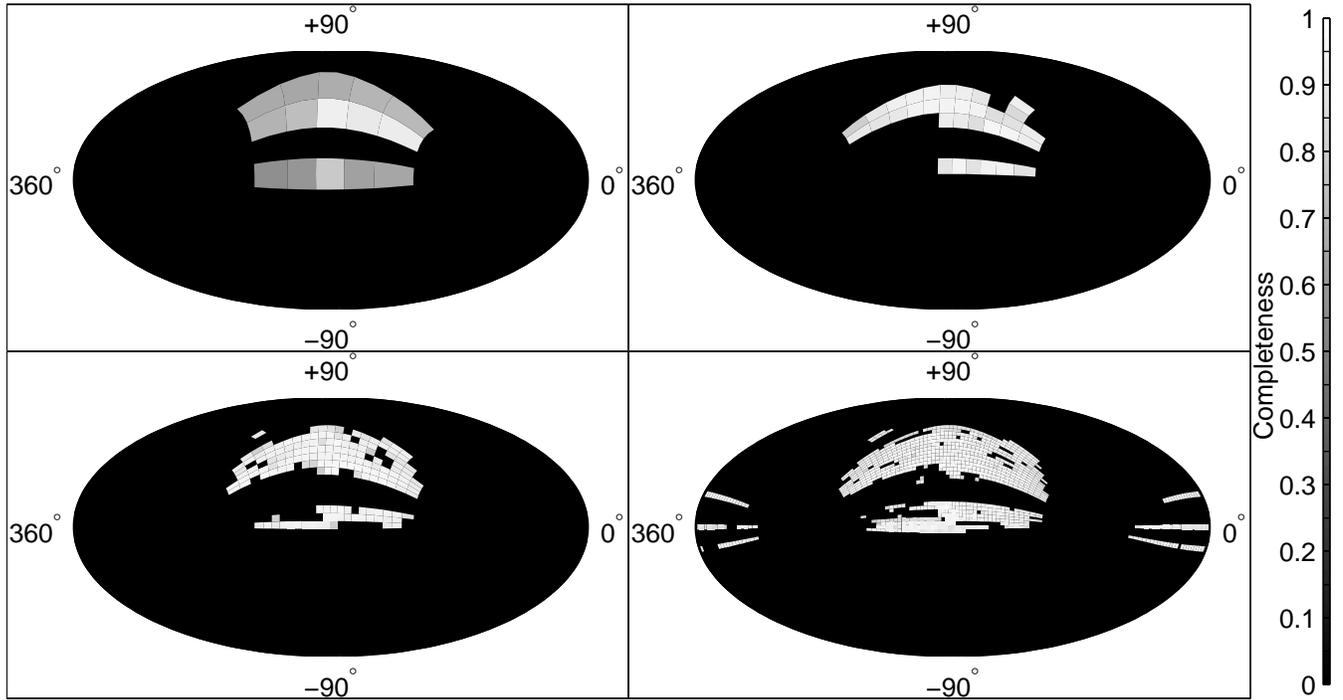}

\caption{\label{fig:angularcells}The SDSS DR5 angular mask pixelized at the
four different resolutions used to partition the survey into cells,
shown in Hammer-Aitoff projection in equatorial co-ordinates. Shading
indicates completeness level: 0 per cent is black, 100 per cent is
white.}
\end{figure*}

To create our cells, we first divide the sky into two-dimensional
`pixels' at four different angular resolutions using the SDSSPix pixelization
scheme%
\footnote{See http://lahmu.phyast.pitt.edu/~scranton/SDSSPix/%
} as implemented by an updated version of the angular mask processing
software \noun{mangle} \citep{2004MNRAS.349..115H,2007arXiv0711.4352S}.
The angular selection function $\bar{n}\left(\hat{\bmath{r}}\right)$
is averaged over each pixel to obtain the completeness. To reduce
the effects of pixels on the edge of the survey area or in regions
affected by internal holes in the survey, we apply a cut on pixel
completeness: we only use pixels with a completeness higher than 80
per cent (50 per cent for the lowest angular resolution). Figure~\ref{fig:angularcells}
shows the pixelized SDSS angular mask at our four different resolutions,
including only the pixels that pass our completeness cut. The different
angular resolutions have 15, 33, 157, and 901 of these angular pixels
respectively. At the lowest resolution, each pixel covers 353 square
degrees, and the angular area of the pixels decreases by a factor
of $1/4$ at each resolution level, yielding pixels covering 88, 22,
and 5 square degrees at the three higher resolutions.%

To produce three-dimensional cells from our pixels, we divide each
comparison volume into radial shells of equal volume. We choose the
number of radial subdivisions at each angular resolution in each comparison
volume such that our cells are approximately cubical, i.e., the radial
extent of a cell is approximately equal to its transverse (angular)
extent. This procedure makes cells that are not quite perfect cubes
-- there is some slight variation in the cell shapes, with cells on
the near edge of the volume slightly elongated radially and cells
on the far edge slightly flattened. We state all of our results as
a function of cell size $\elll$, defined as the cube root of the
cell volume. At the lowest resolution, there is just 1 radial shell
for each volume; at the next resolution, we have 3 radial shells for
volumes V4 and V5 and 2 radial shells for the other volumes. There
are 5 radial shells at the second highest resolution, and 10 at the
highest.

\begin{figure*}
\includegraphics[width=1\textwidth]{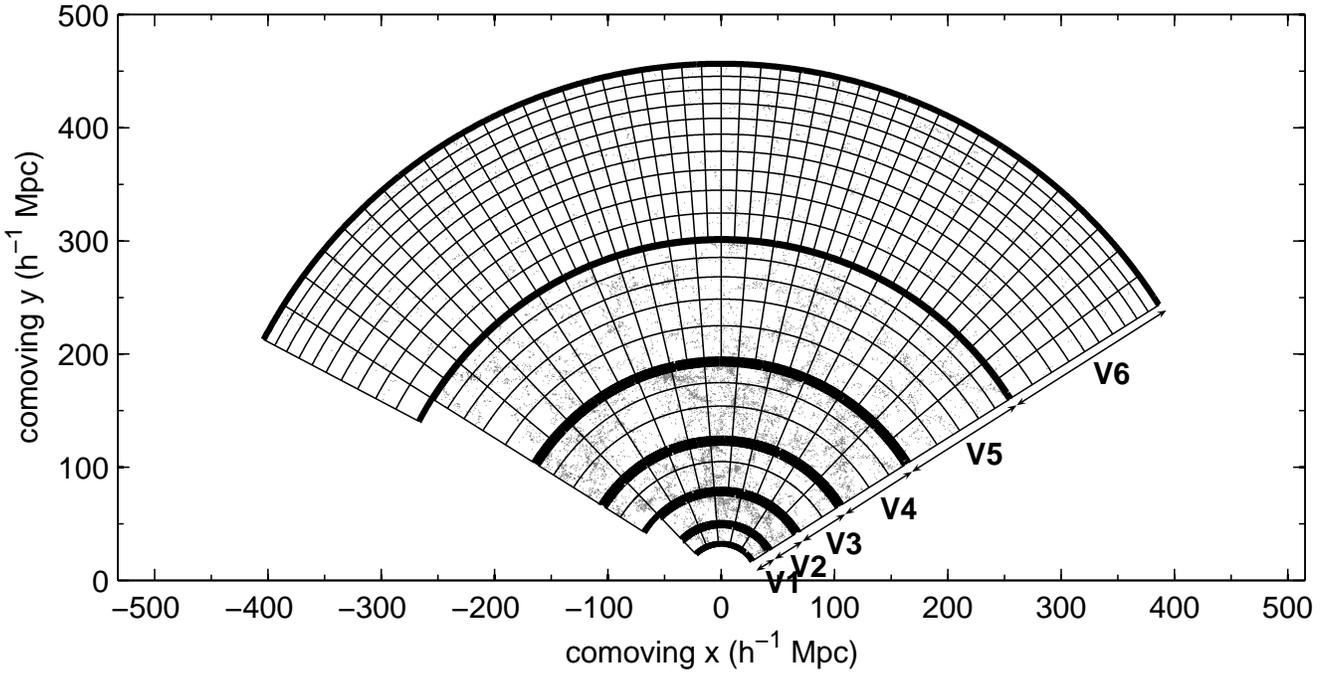}

\caption{\label{fig:radialcells}A radial slice of the SDSS survey volume
divided into cells of size $\sim20{\, h}^{-1}\rmn{Mpc}$ with the
galaxies in each cell (after finger-of-god compression) shown in grey.}
\end{figure*}

Since each comparison volume is at a different distance from us, the
angular geometry gives us cells of different physical size in each
of the volumes. At the lowest resolution, where there is only one
shell in each volume, the cell size is $14\, h^{-1}\rmn{Mpc}$ in
V1 and $134\, h^{-1}\rmn{Mpc}$ in V6. At the highest resolution,
the cell size is $1.7\, h^{-1}\rmn{Mpc}$ in V1 and $16\, h^{-1}\rmn{Mpc}$
in V6. Figure~\ref{fig:radialcells} shows the cells in each volume
V1-V6 that are closest to a size of $\sim20{\, h}^{-1}\rmn{Mpc}$,
the range in which the length scales probed by the different volumes
overlap. (These are the cells used to produce the results shown in
Fig.~\ref{fig:lum-bias}.) %

\subsection{\label{sub:Relative-Bias-Framework}Relative Bias Framework}

Our task is to quantify the relationship between two fractional overdensity
fields $\delta_{1}\left(\bmath{x}\right)\equiv\rho_{1}\left(\bmath{x}\right)/\bar{\rho}_{1}-1$
and $\delta_{2}\left(\bmath{x}\right)\equiv\rho_{2}\left(\bmath{x}\right)/\bar{\rho}_{2}-1$
representing two different types of objects. This framework is commonly
used with types (1,2) representing (dark matter, galaxies), or as
in \citet{2000ApJ...544...63B}, \citet{2005MNRAS.356..247W}, and
\citet{2005MNRAS.356..456C}, (early-type galaxies, late-type galaxies).
Here we use it to represent (more luminous galaxies, dimmer galaxies)
or (red galaxies, blue galaxies) to compare the samples described
in Section~\ref{sub:Overlapping-Volume-Limited-Samples}. Galaxies
are of course discrete objects, and as customary, we use the continuous
field $\rho_{\alpha}\left(\bmath{x}\right)$ (where $\alpha$=1 or
2) to formally refer to the expectation value of the Poisson point
process involved in distributing the type $\alpha$ galaxies.

The simplest (and frequently assumed) relationship between $\delta_{1}$
and $\delta_{2}$ is linear deterministic bias: \begin{equation}
\delta_{2}\left(\bmath{x}\right)=b_{\rmn{lin}}\delta_{1}\left(\bmath{x}\right),\label{eq:linearbias}\end{equation}
 where $b_{\rmn{lin}}$ is a constant parameter. This model cannot
hold in all cases -- note that it can give negative densities if $b_{\rmn{lin}}>1$
-- but is typically a reasonable approximation on cosmologically large
length scales where the density fluctuations $\delta_{1}\ll1$, as
is the case for the measurements of the large scale power spectrum
recently used to constrain cosmological parameters \citep{2004ApJ...606..702T,2005MNRAS.362..505C,2006PhRvD..74l3507T}

More complicated models allow for non-linearity and stochasticity,
as described in detail in \citet{1999ApJ...520...24D}:

\begin{equation}
\delta_{2}\left(\bmath{x}\right)=b\left[\delta_{1}\left(\bmath{x}\right)\right]\delta_{1}\left(\bmath{x}\right)+\epsilon\left(\bmath{x}\right),\label{eq:nonlinearbias}\end{equation}
 where the bias $b$ is now a (typically slightly non-linear) function
of $\delta_{1}$. The stochasticity is represented by a random field
$\epsilon$ -- allowing for stochasticity removes the restriction
implied by deterministic models that the peaks of $\delta_{1}$ and
$\delta_{2}$ must coincide spatially. Stochasticity is basically
the scatter in the relationship between the two density fields due
to physical variables besides the local matter density. Non-local
galaxy formation processes can also give rise to stochasticity, as
discussed in \citet{1999ApJ...525..543M}.

We estimate the overdensity of galaxies of type $\alpha$ in cell
$i$ by

\begin{equation}
g_{\alpha}^{(i)}\equiv\frac{N_{\alpha}^{(i)}-\bar{N}_{\alpha}^{(i)}}{\bar{N}_{\alpha}^{(i)}},\label{eq:countsincells}\end{equation}
 where $N_{\alpha}^{(i)}$ is the number of observed type $\alpha$
galaxies in cell $i$ and $\bar{N}_{\alpha}^{(i)}$ is the expected
number of such galaxies, computed from the average angular selection
function in the pixel and normalized so that the sum of $\bar{N}_{\alpha}^{(i)}$
over all cells in the comparison volume matches the total number of
observed type $\alpha$ galaxies. The $n$-dimensional vectors \begin{equation}
{\bmath{g}}_{\alpha}\equiv\left(\begin{array}{c}
g_{\alpha}^{(1)}\\
\vdots\\
g_{\alpha}^{(n)}\end{array}\right)\label{eq:g_vec1}\end{equation}
 contain the counts-in-cells data to which we apply the statistical
analyses in Sections~\ref{sub:The-Null-buster-Test}~and~\ref{sub:Maximum-Likelihood-Method}.

The covariance matrix of $\bmath{g}$ is given by\begin{equation}
\left\langle g_{\alpha}^{(i)}g_{\beta}^{(j)}\right\rangle =\left\langle \delta_{\alpha}^{(i)}\delta_{\beta}^{(j)}\right\rangle +\delta_{\alpha\beta}{\mathbfss{N}}_{\alpha}^{ij},\label{eq:gdef}\end{equation}
 where $\delta_{\alpha}^{(i)}$ is the average of $\delta_{\alpha}\left(\bmath{x}\right)$
over cell $i$ and, making the customary assumption that the shot
noise is Poissonian, the shot noise covariance matrix ${\mathbfss{N}}_{\alpha}$
is given by \begin{equation}
{\mathbfss{N}}_{\alpha}^{ij}\equiv\delta_{ij}/\bar{N}_{\alpha}^{(i)}.\label{eq:shotnoise}\end{equation}
 %The shot noise is typically assumed to be Poissonian.
For comparing pairs of different types of galaxies, we construct the
data vector \begin{equation}
\bmath{g}\equiv\left(\begin{array}{c}
{\bmath{g}}_{1}\\
{\bmath{g}}_{2}\end{array}\right),\label{eq:g_vec}\end{equation}
 which has a covariance matrix \begin{equation}
\mathbfss{C}\equiv\left\langle \bmath{g}{\bmath{g}}^{T}\right\rangle =\mathbfss{S}+\mathbfss{N},\label{eq:cov_matrix}\end{equation}
 with\begin{equation}
\mathbfss{N}\equiv\left(\begin{array}{cc}
{\mathbfss{N}}_{1} & 0\\
0 & {\mathbfss{N}}_{2}\end{array}\right),\qquad\mathbfss{S}\equiv\left(\begin{array}{cc}
{\mathbfss{S}}_{11} & {\mathbfss{S}}_{12}\\
{\mathbfss{S}}_{12} & {\mathbfss{S}}_{22}\end{array}\right),\label{eq:N_and_S}\end{equation}
 and the elements of the matrix $\mathbfss{S}$ given by

%\begin{eqnarray}
%{\mathbfss{S}}_{11}^{ij} & = & \left\langle \delta_{1}^{(i)}\delta_{1}^{(j)}\right\rangle ,\nonumber \\
%{\mathbfss{S}}_{12}^{ij} & = & \left\langle \delta_{1}^{(i)}\delta_{2}^{(j)}\right\rangle ,\nonumber \\
%{\mathbfss{S}}_{22}^{ij} & = & \left\langle \delta_{2}^{(i)}\delta_{2}^{(j)}\right\rangle .\label{eq:s_matrices}\end{eqnarray}
\begin{equation}
{\mathbfss{S}}_{\alpha\beta}^{ij}=\left\langle \delta_{\alpha}^{(i)}\delta_{\beta}^{(j)}\right\rangle .\label{eq:s_matrices}\end{equation}
 The diagonal form of $\mathbfss{N}$ in equation~\eqref{eq:N_and_S}
assumes that there are no correlations between the shot noise of type
1 and type 2 galaxies within a given cell $i$ -- this means that
the two galaxy distributions are treated as independent Poisson processes
that sample related density distributions $\delta_{1}\left(\bmath{x}\right)$
and $\delta_{2}\left(\bmath{x}\right)$. Although one might expect
the fact that the counts of type 1 and type 2 galaxies in a cell is
constrained to be equal to the total number of galaxies in the cell
could induce correlations in the shot noise, we do not explicitly
use the combined total count in our analyses -- uncorrelated shot
noise is thus a reasonable assumption.

Regarding the matrix $\mathbfss{S}$, other counts-in-cells analyses
often assume that the correlations between different cells can be
ignored, i.e., $\left\langle \delta_{\alpha}^{(i)}\delta_{\beta}^{(j)}\right\rangle =0$
unless $i=j$. Here we account for cosmological correlations by computing
the elements of $\mathbfss{S}$ using the best-fitting $\Lambda$CDM
matter power spectrum as we will now describe in detail. The power
spectrum $P_{\alpha\beta}\left(\bmath{k}\right)$ is defined as $\left\langle \hat{\delta}_{\alpha}\left(\bmath{k}\right)\hat{\delta}_{\beta}\left({\bmath{k}}^{\prime}\right)^{\dagger}\right\rangle =\left(2\pi\right)^{3}\delta^{D}\left(\bmath{k}-{\bmath{k}}^{\prime}\right)P_{\alpha\beta}\left(\bmath{k}\right)$,
where $\hat{\delta}_{\alpha}\left(\bmath{k}\right)\equiv\int e^{-i\bmath{k}\cdot\bmath{x}}\delta_{\alpha}\left(\bmath{x}\right)d^{3}\bmath{x}$
is the Fourier transform of the overdensity field. $P_{11}\left(\bmath{k}\right)$
and $P_{22}\left(\bmath{k}\right)$ are the power spectra of type
1 and 2 galaxies respectively, and $P_{12}\left(\bmath{k}\right)$
is the cross spectrum between type 1 and 2 galaxies. We assume isotropy
and homogeneity, so that $P_{\alpha\beta}\left(\bmath{k}\right)$
is a function only of $k\equiv\left|\bmath{k}\right|$, and rewrite
the galaxy power spectra in terms of the matter power spectrum $P\left(k\right)$:

\begin{eqnarray}
P_{11}\left(k\right) & = & b_{1}\left(k\right)^{2}P\left(k\right)\nonumber \\
P_{12}\left(k\right) & = & {b_{1}\left(k\right)b}_{2}\left(k\right)r_{12}\left(k\right)P\left(k\right)\nonumber \\
P_{22}\left(k\right) & = & b_{2}\left(k\right)^{2}P\left(k\right),\label{eq:Pks}\end{eqnarray}
 which defines the functions $b_{1}\left(k\right)$, $b_{2}\left(k\right)$,
and $r_{12}\left(k\right)$.

To calculate $\left\langle \delta_{\alpha}^{(i)}\delta_{\beta}^{(j)}\right\rangle $
exactly, we need to convolve $\delta_{\alpha}(\bmath{x})$ with a
filter representing cell $i$ and $\delta_{\beta}(\bmath{x})$ with
a filter representing cell $j$. This is complicated since our cells,
while all roughly cubical, have slightly different shapes. We therefore
use an approximation of a spherical top hat smoothing filter with
radius $R$: $w\left(r,R\right)\equiv3/(4\pi R^{3})\Theta(R-r)$ with
the Fourier transform given by \begin{equation}
\hat{w}\left(k,R\right)=\frac{3}{\left(kR\right)^{3}}\left[\sin\left(kR\right)-kR\cos\left(kR\right)\right].\end{equation}
 $R$ is chosen so that the effective scale corresponds to cubes with
side length $\elll$: $R=\sqrt{5/12}\elll$, where $\elll$ is the
cell size defined in Section~\ref{sub:Counts-in-Cells-Methodology}.
(See p. 500 in \citealt{1999coph.book.....P} for derivation of the
$\sqrt{5/12}$ factor.) This gives

\begin{equation}
\left\langle \delta_{\alpha}^{(i)}\delta_{\beta}^{(j)}\right\rangle =\frac{1}{2\pi^{2}}\int_{0}^{\infty}\frac{\sin\left(kr_{ij}\right)}{kr_{ij}}P_{\alpha\beta}\left(k\right)\left|\hat{w}\left(k,R\right)\right|^{2}k^{2}{\rm {d}k,}\label{eq:s_elements}\end{equation}
 where $r_{ij}$ is the distance between the centres of cells $i$
and $j$. The kernel of this integrand -- meaning everything besides
$P_{\alpha\beta}\left(k\right)$ here -- typically peaks at $k\sim1/R$
and is only non-negligible in a range of $\Delta\log_{10}k\sim1$.
Assuming that the functions $b_{1}\left(k\right)$, $b_{2}\left(k\right)$,
and $r_{12}\left(k\right)$ vary slowly with $k$ over this range,
they can be approximated by their values at $k_{\rmn{peak}}\equiv1/R=\sqrt{12/5}/\elll$
and pulled outside the integral, allowing us to write

\begin{equation}
\mathbfss{S}=\sigma_{1}^{2}\left(\elll\right)\left[\begin{array}{cc}
{\mathbfss{S}}_{M} & b_{\rmn{rel}}\left(\elll\right)r_{\rmn{rel}}\left(\elll\right){\mathbfss{S}}_{M}\\
b_{\rmn{rel}}\left(\elll\right)r_{\rmn{rel}}\left(\elll\right){\mathbfss{S}}_{M} & b_{\rmn{rel}}\left(\elll\right)^{2}{\mathbfss{S}}_{M}\end{array}\right]\label{eq:smatrix}\end{equation}
 where $\sigma_{1}^{2}\left(\elll\right)\equiv b_{1}\left(k_{\rmn{peak}}\right)^{2}$,
$b_{\rmn{rel}}\left(\elll\right)\equiv b_{2}\left(k_{\rmn{peak}}\right)/b_{1}\left(k_{\rmn{peak}}\right)$,
$r_{\rmn{rel}}\left(\elll\right)\equiv r_{12}\left(k_{\rmn{peak}}\right)$,
and ${\mathbfss{S}}_{M}$ is the correlation matrix for the underlying
matter density:

\begin{equation}
{\mathbfss{S}}_{M}^{ij}=\frac{1}{2\pi^{2}}\int_{0}^{\infty}\frac{\sin\left(kr_{ij}\right)}{kr_{ij}}P\left(k\right)\left|\hat{w}\left(k,R\right)\right|^{2}k^{2}{\rm {d}k.}\label{eq:sd_elements}\end{equation}
 For the matter power spectrum $P\left(k\right)$, we use the fitting
formula from \citet{1999A&A...347..799N} with the best-fitting `vanilla'
parameters from \citet{2004PhRvD..69j3501T} and apply the non-linear
transformation of \citet{2003MNRAS.341.1311S}.

Our primary parameters are the relative bias factor $b_{\rmn{rel}}\left(\elll\right)$,
the relative cross-correlation coefficient $r_{\rmn{rel}}\left(\elll\right)$,
and the overall normalization $\sigma_{1}^{2}\left(\elll\right)$.
The only assumptions we have made in defining these parameters are
homogeneity, isotropy, and that $b_{1}\left(k\right)$, $b_{2}\left(k\right)$,
and $r_{12}\left(k\right)$ vary slowly in $k$. %How do these relate
These parameters are closely related to those in the biasing models
specified in equations~\eqref{eq:linearbias}~and~\eqref{eq:nonlinearbias}:
If linear deterministic biasing holds, then $b_{\rmn{rel}}=b_{\rmn{lin}}$
and $r_{\rmn{rel}}=1$, % {[}any constraints on scale dependence?]
and the addition of either non-linearity or stochasticity will give
$r_{\rmn{rel}}<1$. %{[}implications of scale dependence?]
As discussed in \citet{1999ApJ...525..543M}, stochasticity is expected
to vanish in Fourier space (i.e., $r_{12}\left(k\right)=1$) on large
scales where the density fluctuations are small, but scale dependence
of $b_{1}\left(k\right)$ and $b_{2}\left(k\right)$ can still give
rise to stochasticity in real space. We will measure the parameters
$b_{\rmn{rel}}\left(\elll\right)$ and $r_{\rmn{rel}}\left(\elll\right)$
as a function of scale, thus testing whether the bias is scale dependent
and determining the range of scales on which linear biasing holds.

\subsection{\label{sub:The-Null-buster-Test}The Null-buster Test}

Can the relative bias between dim and luminous galaxies or between
red and blue galaxies be explained by simple linear deterministic
biasing? To address this question, we use the so-called null-buster
test described in \citet{1999ApJ...518L..69T}. For a pair of different
types of galaxies, we calculate a difference map \begin{equation}
\Delta\bmath{g}\equiv{\bmath{g}}_{2}-f{\bmath{g}}_{1}\label{eq:diffmap}\end{equation}
 for a range of values of $f$. If equation~\eqref{eq:linearbias}
holds and $f=b_{\rmn{lin}}$, then the density fluctuations cancel
and $\Delta\bmath{g}$ will contain only shot noise, with a covariance
matrix $\left\langle \Delta\bmath{g}\Delta{\bmath{g}}^{T}\right\rangle ={\mathbfss{N}}_{\Delta}\equiv{\mathbfss{N}}_{2}+f^{2}{\mathbfss{N}}_{1}$
-- this is our null hypothesis.

If equation~\eqref{eq:linearbias} does not hold and the covariance
matrix is instead given by $\left\langle \Delta\bmath{g}\Delta{\bmath{g}}^{T}\right\rangle ={\mathbfss{N}}_{\Delta}+{\mathbfss{S}}_{\Delta}$,
where ${\mathbfss{S}}_{\Delta}$ is some residual signal, then the
most powerful test for ruling out the null hypothesis is the generalized
$\chi^{2}$ statistic \citep{1998ApJ...500L..79T} \begin{equation}
\nu\equiv\frac{\Delta{\bmath{g}}^{T}{\mathbfss{N}}_{\Delta}^{-1}{\mathbfss{S}}_{\Delta}{\mathbfss{N}}_{\Delta}^{-1}\Delta\bmath{g}-\rmn{Tr}\left({\mathbfss{N}}_{\Delta}^{-1}{\mathbfss{S}}_{\Delta}\right)}{\left[\rmn{2\, Tr}\left({\mathbfss{N}}_{\Delta}^{-1}{\mathbfss{S}}_{\Delta}{\mathbfss{N}}_{\Delta}^{-1}{\mathbfss{S}}_{\Delta}\right)\right]^{1/2}},\label{eq:nullbuster}\end{equation}
 which can be interpreted as the significance level (i.e. the number
of `sigmas') at which we can rule out the null hypothesis. As detailed
in \citet{1999ApJ...519..513T}, this test assumes that the Poissonian
shot noise contribution can be approximated as Gaussian but makes
no other assumptions about the probability distribution of $\Delta\bmath{g}$.
It is a valid test for any choice of ${\mathbfss{S}}_{\Delta}$ and
reduces to a standard $\chi^{2}$ test if ${\mathbfss{S}}_{\Delta}={\mathbfss{N}}_{\Delta}$,
but it rules out the null hypothesis with maximum significance in
the case where ${\mathbfss{S}}_{\Delta}$ is the true residual signal.

Using equations~\eqref{eq:N_and_S},~\eqref{eq:s_matrices},~and~\eqref{eq:smatrix},
the covariance matrix of $\Delta\bmath{g}$ can be written as \begin{equation}
\left\langle \Delta\bmath{g}\Delta{\bmath{g}}^{T}\right\rangle =\sigma_{1}^{2}\left(f^{2}-2b_{{\rm \rmn{rel}}}r_{\rmn{rel}}f+b_{\rmn{rel}}^{2}\right){\mathbfss{S}}_{M}+{\mathbfss{N}}_{\Delta},\end{equation}
 where ${\mathbfss{S}}_{M}$ is given by equation~\eqref{eq:sd_elements}.
We use ${\mathbfss{S}}_{\Delta}={\mathbfss{S}}_{M}$ in equation~\eqref{eq:nullbuster}
(note that $\nu$ is independent of the normalization of $\mathbfss{S}$,
which scales out) since deviations from linear deterministic bias
are likely to be correlated with large-scale structure. %The residual signal is minimized when $f=b_{{\rm \rmn{rel}}}r_{\rmn{rel}}$,
%{[}or is it? calculating $\Delta\bmath g_{flipped}\equiv\bmath g_{1}-f_{flipped}\bmath g_{2}$
%seems like it should be minimized at $f_{flipped}=r_{\rmn{rel}}/b_{\rmn{rel}}$,
%but if I do this I get $\sqrt{ff_{flipped}}=r_{\rmn{rel}}=1$ even
%when the test says the null hypothesis is ruled out.] and if $r_{\rmn{rel}}=1$
%the residual signal cancels out completely.

To apply the null-buster test, we compute $\nu$ as a function of
$f$ and then minimize it. If the minimum value $\nu_{\rmn{min}}>2$,
we rule out linear deterministic bias at $>2\sigma$. If the null
hypothesis cannot be ruled out and we choose to accept it as an accurate
description of the data, we can use the value of $f$ that gives $\nu_{\rmn{min}}$
as a measure of $b_{\rmn{rel}}$. 

We calculate the uncertainty on $b_{\rmn{rel}}$ using two different
methods. The first method makes use of the fact that $\nu$ is generalized
$\chi^{2}$ statistic: the uncertainty on $b_{\rmn{rel}}$, is given
by the range in $f$ that gives $\sqrt{2N}\left(\nu-\nu_{\rmn{min}}\right)\le1$,
where $N$ is the number of degrees of freedom (equal to the number
of cells minus 1 fitted parameter). This is a generalization of the
standard $\Delta\chi^{2}=1$ uncertainty since $\nu$ is a generalization
of $\left(\chi^{2}-N\right)/\sqrt{2N}$. The second method uses jackknife
resampling, which is described in Section~\ref{sub:Jackknifes} along
with a comparison of the two methods. We present all of our results
derived from the null-buster test using the uncertainties from the
jackknife method.

\subsection{\label{sub:Maximum-Likelihood-Method}Maximum Likelihood Method}

In addition to the null-buster test, we use a maximum likelihood analysis
to determine the parameters $b_{\rmn{rel}}$ and $r_{\rmn{rel}}$.
Our method is a generalization of the maximum likelihood method used
in previous papers, accounting for correlations between different
cells but making a somewhat different set of assumptions.

In \citet{2000ApJ...544...63B}, \citet{2005MNRAS.356..247W}, \citet{2005MNRAS.356..456C},
the probability of observing $N_{1}$ galaxies of type 1 and $N_{2}$
galaxies of type 2 in a given cell is expressed as

\begin{eqnarray}
\lefteqn{P\left(N_{1},N_{2}\right)=\int_{-1}^{\infty}\int_{-1}^{\infty}\rmn{Poiss}\left[N_{1},\bar{N}_{1}\left(1+\delta_{1}\right)\right]}\nonumber \\
 &  & \times\rmn{Poiss}\left[N_{2},\bar{N}_{2}\left(1+\delta_{2}\right)\right]f\left({\delta_{1},\delta}_{2},\alpha\right)\rmn{d}\delta_{1}\rmn{d}\delta_{2},\label{eq:wild}\end{eqnarray}
 where $f\left({\delta_{1},\delta}_{2},\alpha\right)$ is the joint
probability distribution of $\delta_{1}$ and $\delta_{2}$ in one
cell, $\alpha$ represents a set of parameters which depend on the
biasing model, and $\rmn{Poiss}\left(N,\lambda\right)\equiv\lambda^{N}e^{-\lambda}/N!$
is the Poisson probability to observe $N$ objects given a mean value
$\lambda$. The likelihood function for $n$ cells is then given by
\begin{equation}
\mathcal{L}\left(\alpha\right)\equiv\prod_{i=1}^{n}P\left(N_{1}^{(i)},N_{2}^{(i)}\right),\label{eq:uncorr_L}\end{equation}
 which is minimized with respect to the parameters $\alpha$. This
treatment makes two assumptions: it neglects correlations between
different cells and it assumes that the galaxy discreteness is Poissonian.
These assumptions greatly simplify the computation of $\mathcal{L}$,
but are understood to be approximations to the true process. Cosmological
correlations are known to exist on large scales, although their impact
on counts-in-cells analyses has been argued to be small \citep{1995ApJ...438...49B,2005MNRAS.356..456C}.
Semi-analytical modelling \citep{2001MNRAS.322..901S,2002ApJ...575..587B},
$N$-body simulation \citep{2002MNRAS.333..730C,2004ApJ...609...35K},
and smoothed particle hydrodynamic simulation \citep{2003ApJ...593....1B}
investigations suggest that the probability distribution for galaxies/haloes
is sub-Poissonian in some regimes, and in fact non-Poissonian behaviour
is implied by observations as well \citep{2003MNRAS.339.1057Y,2005MNRAS.356..247W}.

Dropping these two assumptions, we can write a more general expression
for the likelihood function for $n$ cells: \begin{eqnarray}
\lefteqn{{\mathcal{L}\left(\alpha,\beta\right)\equiv P\left(N_{1}^{\left(1\right)},\ldots,N_{1}^{\left(n\right)},N_{2}^{\left(1\right)},\ldots,N_{2}^{\left(n\right)}\right)=}}\nonumber \\
 &  & \int_{-1}^{\infty}\ldots\int_{-1}^{\infty}\left[\prod_{i=1}^{n}P_{g}\left(N_{1}^{\left(i\right)},\bar{N}_{1}^{\left(i\right)}\left(1+\delta_{1}^{\left(i\right)}\right),\beta\right)\right]\nonumber \\
 &  & \times\left[\prod_{j=1}^{n}P_{g}\left(N_{2}^{\left(j\right)},\bar{N}_{2}^{\left(j\right)}\left(1+\delta_{2}^{\left(j\right)}\right),\beta\right)\right]\nonumber \\
 &  & \times f\left(\delta_{1}^{\left(1\right)},\ldots,\delta_{1}^{\left(n\right)},\delta_{2}^{\left(1\right)},\ldots,\delta_{2}^{\left(n\right)},\alpha\right)\nonumber \\
 &  & \times\rmn{d}\delta_{1}^{\left(1\right)}\ldots\rmn{d}\delta_{1}^{\left(n\right)}\rmn{d}\delta_{2}^{\left(1\right)}\ldots\rmn{d}\delta_{2}^{\left(n\right)},\label{eq:corr_L}\end{eqnarray}
 where $\rmn{Poiss}\left(N,\lambda\right)$ has been replaced with
a generic probability $P_{g}\left(N,\lambda,\beta\right)$ for the
galaxy distribution parameterized by some parameters $\beta$ and
$f\left(\delta_{1}^{\left(1\right)},\ldots,\delta_{1}^{\left(n\right)},\delta_{2}^{\left(1\right)},\ldots,\delta_{2}^{\left(n\right)},\alpha\right)$
is a joint probability distribution relating $\delta_{1}$ and $\delta_{2}$
in all cells. In practice, this would be prohibitively difficult to
calculate as it involves $2n$ integrations \citep{1997astro.ph.12074D},
and would require a reasonable parameterized form for $P_{g}\left(N,\lambda,\beta\right)$
as well as $f\left(\delta_{1}^{\left(1\right)},\ldots,\delta_{1}^{\left(n\right)},\delta_{2}^{\left(1\right)},\ldots,\delta_{2}^{\left(n\right)},\alpha\right)$.

In this paper, we take a simpler approach and approximate the probability
distribution for our data vector $\bmath{g}$ to be Gaussian with
the covariance matrix $\mathbfss{C}$ as defined by equations~\eqref{eq:cov_matrix},~\eqref{eq:N_and_S}~and~\eqref{eq:smatrix},
and use this to define our likelihood function in terms of the parameters
$\sigma_{1}^{2}$, $b_{\rmn{rel}}$, and $r_{\rmn{rel}}$:

\begin{eqnarray}
\lefteqn{{\mathcal{L}\left(\sigma_{1}^{2},b_{\rmn{rel}},r_{\rmn{rel}}\right)\equiv P\left(g_{1}^{\left(1\right)},\ldots,g_{1}^{\left(n\right)},g_{2}^{\left(1\right)},\ldots,g_{2}^{\left(n\right)}\right)}}\nonumber \\
 &  & =\frac{1}{\left(2\pi\right)^{n}\left|\mathbfss{C}\right|^{1/2}}\exp\left(-\frac{1}{2}{\bmath{g}}^{\dagger}{\mathbfss{C}}^{-1}\bmath{g}\right).\label{eq:likelihood}\end{eqnarray}
 Note that this includes the shot noise since $\mathbfss{C}=\mathbfss{S}+\mathbfss{N}$,
and is not precisely equivalent to assuming that $P_{g}$ and $f$
in equation~\eqref{eq:corr_L} are Gaussian. 

For $r_{\rmn{rel}}$ values of $\left|r_{\rmn{rel}}\right|>1$, the
matrix $\mathbfss{C}$ is singular, and thus the likelihood function
cannot be computed. Hence this analysis method automatically incorporates
the constraint that $\left|r_{\rmn{rel}}\right|\le1$, which is physically
expected for a cross-correlation coefficient.

To determine the best fit values of our parameters for each pairwise
comparison, we maximize $2\ln\mathcal{L}\left(\sigma_{1}^{2},b_{\rmn{rel}},r_{\rmn{rel}}\right)$
with respect to $\sigma_{1}^{2}$, $b_{\rmn{rel}}$, and $r_{\rmn{rel}}$.
Since our method of comparing pairs of galaxy samples primarily probes
the relative biasing between the two types of galaxies, it is not
particularly sensitive to $\sigma_{1}^{2}$, which represents the
bias of type 1 galaxies relative to the dark matter power spectrum
used in equation~\eqref{eq:sd_elements}. Thus we marginalize over
$\sigma_{1}^{2}$ and calculate the uncertainty on $b_{\rmn{rel}}$
and $r_{\rmn{rel}}$ using the $\Delta\left(2\ln\mathcal{L}\right)=1$
contour in the $b_{\rmn{rel}}$-$r_{\rmn{rel}}$ plane. This procedure
is discussed in more detail in Section~\ref{sub:Likelihood-contour-plots}.

\section{Results}

\label{Results}

\subsection{\label{sub:Null-buster-Results}Null-buster Results}

To test the deterministic linear bias model, we apply the null-buster
test described in Section~\ref{sub:The-Null-buster-Test} to the
pairs of galaxy samples described in Section~\ref{sub:Overlapping-Volume-Limited-Samples}.
For studying the luminosity-dependent bias, we use the galaxies in
the more luminous bin as the type 1 galaxies and the dimmer bin as
the type 2 galaxies for each pair of neighbouring luminosity bins,
and repeat this in each volume V1-V6. We do this for all galaxies
and also for red and blue galaxies separately. For the colour dependence,
we use red galaxies as type 1 and blue galaxies as type 2, and again
repeat this in each volume. To determine the scale dependence, we
repeat all of these tests for four different values of the cell size
$\elll$ as described in Section~\ref{sub:Counts-in-Cells-Methodology}.

\subsubsection{\label{sub:linear-deterministic}Is the bias linear and deterministic?}

\begin{figure*}
\includegraphics[width=1\textwidth]{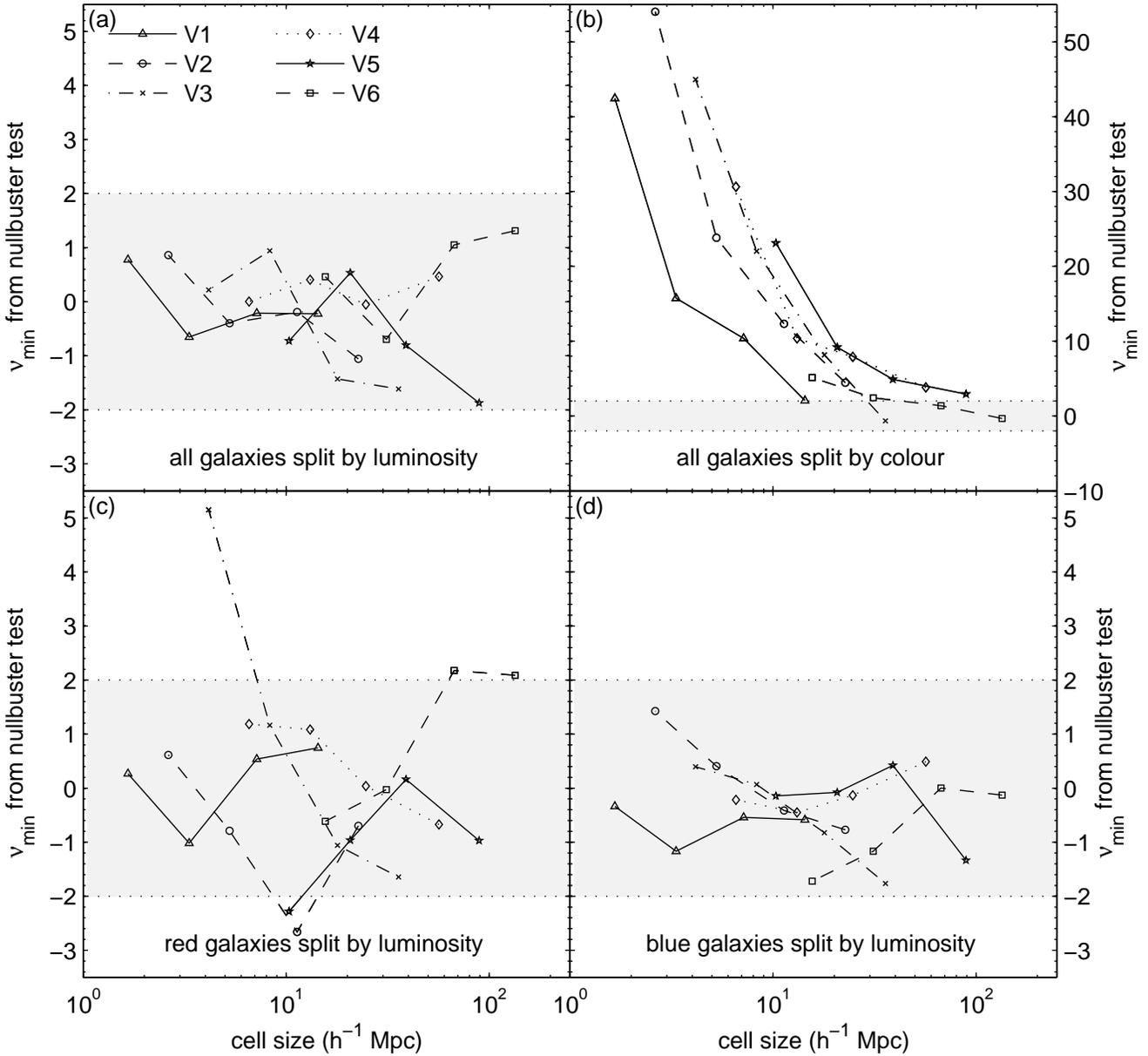}

\caption{\label{fig:nullbusterplot}Null-buster results for pairwise comparisons.
$\nu_{\rmn{min}}$ measures the number of sigmas at which deterministic
linear biasing can be ruled out as a model of relative bias between
the two samples being compared. Shaded areas indicate $\left|\nu_{\rmn{min}}\right|<2$,
where data is consistent with the null hypothesis at the $2\sigma$
level. Four different types of pairwise comparison are illustrated:
(a) luminous vs. dim, (b) red vs. blue, (c) luminous red vs. dim red,
and (d) luminous blue vs. dim blue. The different symbols denote the
different comparison volumes V1-V6. The luminosity-dependent bias
(a, c, d) is consistent with deterministic linear biasing but colour-dependent
bias (b) is not.}
\end{figure*}

The results are plotted in Fig.~\ref{fig:nullbusterplot}, which
shows the minimum value of the null-buster test statistic $\nu_{\rmn{min}}$
vs. cell size $\elll$. According to this test, deterministic linear
biasing is in fact an excellent fit for the luminosity-dependent bias:
nearly all $\nu_{\rmn{min}}$ fall within $\left|\nu_{\rmn{min}}\right|<2$,
indicating consistency with the null hypothesis at the $2\sigma$
level. (There are a few exceptions in the case of the red galaxies,
the largest being $\nu_{\rmn{min}}\sim5$ for the smallest cell size
in V3.) For colour-dependent bias, however, deterministic linear biasing
is ruled out quite strongly, especially at smaller scales.

The cases where the null hypothesis survives are quite noteworthy,
since this implies that essentially all of the large clustering signal
that is present in the data (and is visually apparent in Fig.~\ref{fig:4pairwise})
is common to the two galaxy samples and can be subtracted out. For
example, for the V5 luminosity split at the highest angular resolution
($\elll=10\, h^{-1}\rmn{Mpc}$), clustering signal is detected at
$953\sigma$ in the faint sample ($\nu(f)\approx953$ for $f=0$)
and at $255\sigma$ in the bright sample ($\nu(f)\approx255$ for
$f=\infty$), yet the weighted difference of the two maps is consistent
with mere shot noise ($\nu(0.88)\approx-0.63$). This also shows that
no luminosity-related systematic errors afflict the sample selection
even at that low level.%

\subsubsection{\label{sub:scale-dependent}Is the bias independent of scale?}

\begin{figure*}
\includegraphics[width=1\textwidth]{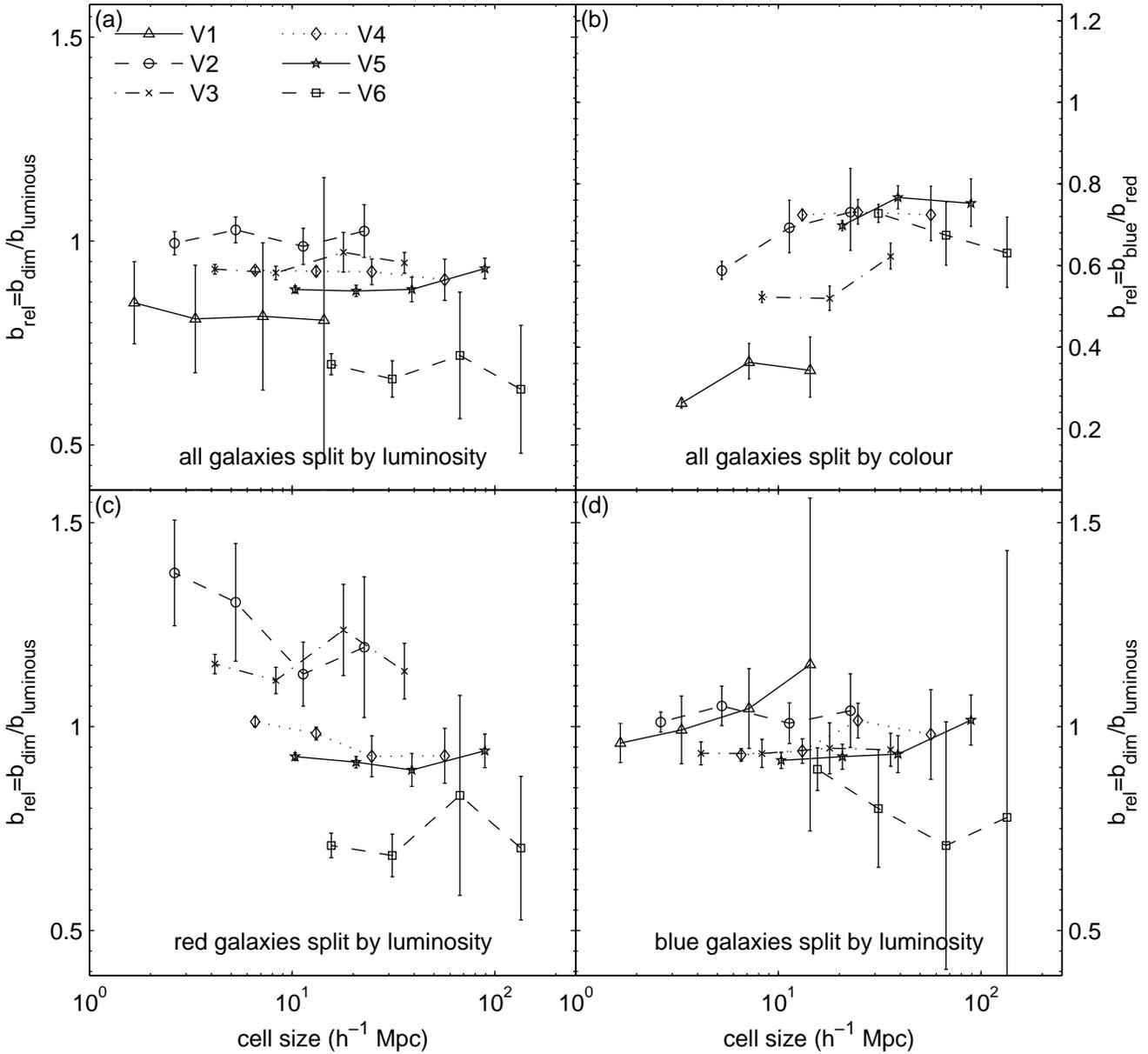}

\caption{\label{fig:biasplot}Relative bias $b_{\rmn{rel}}$ between pairwise
samples. (a) luminous vs. dim, (b) red vs. blue, (c) luminous red
vs. dim red, and (d) luminous blue vs. dim blue, revealing no significant
scale dependence of luminosity- or colour-dependent bias. The $b_{\rmn{rel}}$
values shown for luminosity dependent splittings (a), (c), and (d)
were computed with the null-buster analysis, those shown for the colour-dependent
splitting (b) were computed with the likelihood analysis. The different
symbols denote the different comparison volumes V1-V6.}
\end{figure*}

For the luminosity-dependent bias, we use the value of $f$ that gives
$\nu_{\rmn{min}}$ as a measure of $b_{\rmn{rel}}$, the relative
bias between two neighbouring luminosity bins. Since deterministic
linear bias is ruled out in the case of the colour-dependent bias,
we instead use the value of $b_{\rmn{rel}}$ from the likelihood analysis
here. We find that although the value of $b_{\rmn{rel}}$ depends
on luminosity, it does not appear to depend strongly on scale, as
can be seen in Fig.~\ref{fig:biasplot}: in all plots the curves
appear roughly horizontal. To test this `chi-by-eye' inference of
scale independence quantitatively, we applied a simple $\chi^{2}$
fit on the four data points (or three in the colour-dependent case)
in each volume using a one-parameter model: a horizontal line with
a constant value of $b_{\mathrm{rel}}$. For this fit we use covariance
matrices derived from jackknife resampling, as discussed in Section~\ref{sub:Jackknifes}.
We define this model to be a good fit if the goodness-of-fit value
(the probability that a $\chi^{2}$ as poor as as the value calculated
should occur by chance, as defined in \citealt{1992nrca.book.....P})
exceeds 0.01.%, roughly corresponding to a 99 per cent confidence level. 

We find that this model of no scale dependence is a good fit for all
data sets plotted in Fig.~\ref{fig:biasplot}. We therefore find
no evidence that the luminosity- or colour-dependent bias is scale-dependent
on the scales we probe here$\left(2-160\, h^{-1}\rmn{Mpc}\right)$.
% http://www.nature.com/nature/journal/v257/n5524/abs/257294a0.html
% pegs Coma around 6 Mpc/h$,
This implies that recent cosmological parameter analyses which use
only measurements on scales $\ga60{\, h}^{-1}\rmn{Mpc}$ (e.g., \citealt{2006MNRAS.366..189S,2006PhRvD..74l3507T,2007ApJS..170..377S})
are probably justified in assuming scale independence of luminosity-dependent
bias. %

In comparison to previous work \citep{2005ApJ...630....1Z,2006MNRAS.368...21L},
it is perhaps surprising to see as little scale dependence as we do
-- \citet{2006MNRAS.368...21L} find the luminosity-dependent bias
to vary with scale (see their fig.~4), in contrast to what we find
here. The measurement of luminosity-dependent bias in \citet{2005ApJ...630....1Z}
agrees more closely with our observation of scale independence, but
their their fig.~10 indicates that we might expect to see scale dependence
of the luminosity-dependent bias in the most luminous samples. However,
we measure the bias in our most luminous samples (in V6) at $16-134\, h^{-1}\rmn{Mpc}$,
well above the range probed in \citet{2005ApJ...630....1Z}, so there
is no direct conflict here. Additionally, fig.~13 of \citet{2005ApJ...630....1Z}
and fig.~10 of \citet{2006MNRAS.368...21L} show that correlation
functions of red and blue galaxies have significantly different slopes,
implying that the colour-dependent bias should be strongly scale-dependent
on $0.1-10\, h^{-1}\rmn{Mpc}$ scales. However, the points $>1\, h^{-1}\rmn{Mpc}$
in these plots (the range comparable to the scales we probe here)
do not appear strongly scale dependent, so our results are not inconsistent
with these correlation function measurements. This interpretation
is further supported by recent work \citep{2007ApJ...664..608W} that
finds correlation functions for different luminosities and colours
to be roughly parallel above $\sim1\, h^{-1}\rmn{Mpc}$.

\subsubsection{\label{sub:luminosity-dependent}How bias depends on luminosity}

Our next step is to calculate the relative bias parameter $b/b_{*}$
(the bias relative to $L_{*}$ galaxies) as a function of luminosity
by combining the measured values of $b_{\rmn{rel}}$ between the different
pairs of luminosity bins. This function has been measured previously
using SDSS power spectra \citep{2004ApJ...606..702T} at length scales
of $\sim60\, h^{-1}\rmn{Mpc}$ as well as SDSS \citep{2005ApJ...630....1Z,2006MNRAS.368...21L,2007ApJ...664..608W}
and 2dFGRS \citep{2001MNRAS.328...64N} correlation functions at length
scales of $\sim1\, h^{-1}\rmn{Mpc}$ -- here we measure it at length
scales of $\sim20\, h^{-1}\rmn{Mpc}$. % with smaller error bars by eliminating sample variance.

The bias of each luminosity bin relative to the central bin L4 is
given by

\begin{eqnarray}
 &  & \frac{b_{1}}{b_{4}}=b_{12}b_{23}b_{34},\qquad\frac{b_{2}}{b_{4}}=b_{23}b_{34},\qquad\frac{b_{3}}{b_{4}}=b_{34},\nonumber \\
 &  & \frac{b_{7}}{b_{4}}=\frac{1}{b_{45}b_{56}b_{67}},\qquad\frac{b_{6}}{b_{4}}=\frac{1}{b_{45}b_{56}},\qquad\frac{b_{5}}{b_{4}}=\frac{1}{b_{45}},\label{eq:lum_bias}\end{eqnarray}
 where $b_{\alpha\beta}$ denotes the measured value of $b_{\rmn{rel}}$
between luminosity bins L$\alpha$ and L$\beta$ using all galaxies
and $b_{\alpha}$ denotes the bias of galaxies in luminosity bin L$\alpha$
relative to the dark matter. For each pairwise comparison, we choose
the value of $b_{\rmn{rel}}$ calculated at the resolution where the
cell size is closest to $20\, h^{-1}\rmn{Mpc}$, as illustrated in
Fig.~\ref{fig:radialcells}. (Since we see no evidence for scale
dependence of $b_{\rmn{rel}}$ for the luminosity-dependent bias,
this choice does not strongly influence the results.)

To compute the error bars on $b_{\alpha}/b_{4}$, we rewrite equation~\eqref{eq:lum_bias}
as a linear matrix equation using the logs of the bias values:

\begin{equation}
\left(\begin{array}{cccccc}
1 & -1 & 0 & 0 & 0 & 0\\
0 & 1 & -1 & 0 & 0 & 0\\
0 & 0 & 1 & 0 & 0 & 0\\
0 & 0 & 0 & -1 & 0 & 0\\
0 & 0 & 0 & 1 & -1 & 0\\
0 & 0 & 0 & 0 & 1 & -1\end{array}\right)\left(\begin{array}{c}
\log{b_{1}/b_{4}}\\
\log{b_{2}/b_{4}}\\
\log{b_{3}/b_{4}}\\
\log{b_{5}/b_{4}}\\
\log{b_{6}/b_{4}}\\
\log{b_{7}/b_{4}}\end{array}\right)=\left(\begin{array}{c}
\log{b_{12}}\\
\log{b_{23}}\\
\log{b_{34}}\\
\log{b_{45}}\\
\log{b_{56}}\\
\log{b_{67}}\end{array}\right),\label{eq:lum_matrix}\end{equation}
 or $\mathbfss{A}{\bmath{b}}_{\rmn{log}}={\bmath{b}}_{\rmn{log,rel}}$,
where ${\bmath{b}}_{\rmn{log,rel}}$ is a vector of the log of our
relative bias measurements $b_{\alpha\beta}$, ${\bmath{b}}_{\rmn{log}}$
is a vector of the log of the bias values $b_{\alpha}/b_{4}$, and
$\mathbfss{A}$ is the matrix relating them. We determine the covariance
matrix $\mathbf{\Sigma}_{\rmn{rel}}$ of ${\bmath{b}}_{\rmn{rel}}$
(a vector of the relative bias measurements $b_{\alpha\beta}$) from
the jackknife resampling described in Section~\ref{sub:Jackknifes},
and then compute the covariance matrix $\mathbf{\Sigma}_{\rmn{log,rel}}$
of ${\bmath{b}}_{\rmn{log,rel}}$ by

\begin{equation}
\mathbf{\Sigma}_{\rmn{log,rel}}=\left({\mathbfss{B}}_{\rmn{rel}}^{T}\right)^{-1}\mathbf{\Sigma}_{\rmn{rel}}{\mathbfss{B}}_{\rmn{rel}}^{-1}\label{eq:cov_logrel}\end{equation}
 where ${\mathbfss{B}}_{\rmn{rel}}\equiv\rmn{diag}\left({\bmath{b}}_{\rmn{rel}}\right)$.
We invert equation~\eqref{eq:lum_matrix} to give ${\bmath{b}}_{\rmn{log}}$:

\begin{equation}
{\bmath{b}}_{\rmn{log}}={\mathbfss{A}}^{-1}{\bmath{b}}_{\rmn{log,rel}},\label{eq:invert}\end{equation}
 with the covariance matrix for ${\bmath{b}}_{\rmn{log}}$ given by

\begin{equation}
\mathbf{\Sigma}_{\rmn{log}}=\left({\mathbfss{A}}^{T}\mathbf{\Sigma}_{\rmn{log,rel}}^{-1}\mathbfss{A}\right)^{-1}.\label{eq:cov_log}\end{equation}

We then fit our data with the model used in \citet{2001MNRAS.328...64N}:
$b\left(M\right)/b_{*}=a_{1}+a_{2}\left(L/L_{*}\right)$, parameterized
by $\bmath{a}\equiv\left(a_{1},\, a_{2}\right)$. Here $M$ is the
central absolute magnitude of the bin, $L$ is the corresponding luminosity,
and $M_{*}=-20.83$. We use a weighted least-squares fit that is linear
in the parameters $\left(a_{1},\, a_{2}\right)$ -- that is, we solve
the matrix equation\begin{equation}
\left(\begin{array}{c}
b_{1}/b_{4}\\
b_{2}/b_{4}\\
b_{3}/b_{4}\\
b_{4}/b_{4}\\
b_{5}/b_{4}\\
b_{6}/b_{4}\end{array}\right)=\left(\begin{array}{cc}
1 & L_{1}/L_{4}\\
1 & L_{2}/L_{4}\\
1 & L_{3}/L_{4}\\
1 & L_{4}/L_{4}\\
1 & L_{5}/L_{4}\\
1 & L_{6}/L_{4}\end{array}\right)\left(\begin{array}{c}
a_{1}\\
a_{2}\end{array}\right),\label{eq:model_matrix}\end{equation}
 or $\bmath{b}=\mathbfss{X}\bmath{a}$, where $\bmath{b}$ is a vector
of the bias values $b_{\alpha}/b_{4}$ and $\mathbfss{X}$ is the
matrix representing our model. We solve for $\bmath{a}$ using

\begin{equation}
\bmath{a}=\left({\mathbfss{X}}^{T}\mathbf{\Sigma}^{-1}\mathbfss{X}\right)^{-1}{\mathbfss{X}}^{T}\mathbf{\Sigma}^{-1}\bmath{b}.\label{eq:best_fit_LS}\end{equation}
 Here $\mathbf{\Sigma}$ is the covariance matrix of $\bmath{b}$,
given by\begin{equation}
\mathbf{\Sigma}=\mathbfss{B}\mathbf{\Sigma}_{\rmn{log}}{\mathbfss{B}}^{T},\label{eq:cov_notlog}\end{equation}
 where $\mathbf{\Sigma}_{\rmn{log}}$ is given by equation~\eqref{eq:cov_log}
and $\mathbfss{B}\equiv\rmn{diag}\left(\bmath{b}\right)$. This procedure
gives us the best-fitting values for the parameters $a_{1}$ and $a_{2}$,
accounting for the correlations between the data points that are induced
we compute the bias values $\bmath{b}$ from our relative bias measurements.
We then normalize the model such that $b\left(M_{*}\right)/b_{*}=1$.%
\begin{figure}
\includegraphics[width=1\columnwidth]{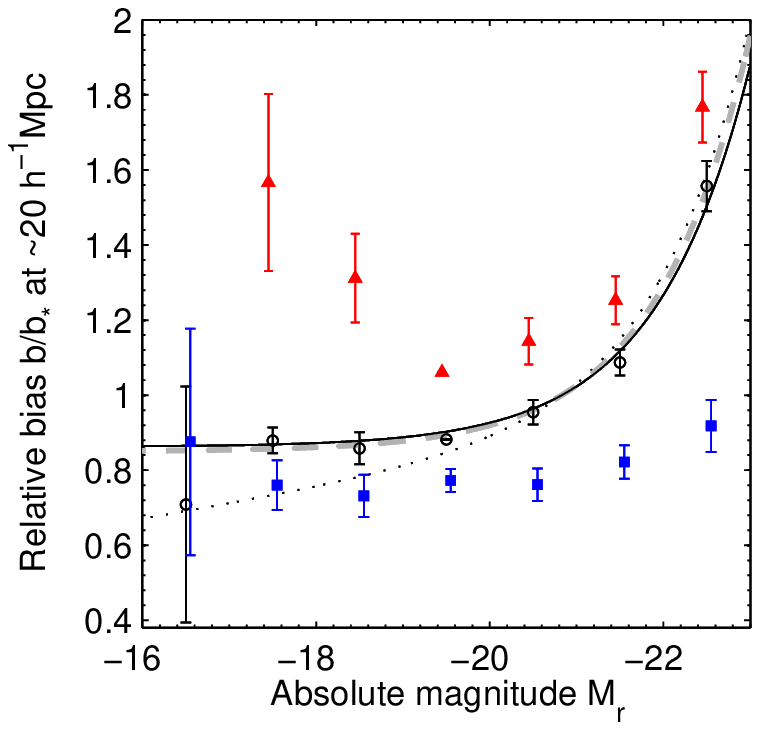}

\caption{\label{fig:lum-bias}Luminosity dependence of bias for all (open
circles), red (solid triangles), and blue (solid squares) galaxies
at a cell size of $\sim20\, h^{-1}\rmn{Mpc}$ from null-buster results.
The solid line is a model fit to the all-galaxy data points, the dotted
line shows the model from \citet{2004ApJ...606..702T}, and the grey
dashed line shows the model from \citet{2001MNRAS.328...64N}. The
\citet{2001MNRAS.328...64N} model has been computed using the SDSS
$r$-band value of $M_{*}=-20.83$.}
\end{figure}

Figure~\ref{fig:lum-bias} shows a plot of $b/b_{*}$ vs. $M$: results
for all galaxies are plotted with black open circles, our best-fitting
model is shown by the solid line, the best-fitting model from \citet{2001MNRAS.328...64N}
is shown by the grey dashed line,and the best-fitting model from \citet{2004ApJ...606..702T}
is shown by the dotted line. The error bars represent the diagonal
elements of $\mathbf{\Sigma}$ from equation~\eqref{eq:cov_notlog}.
Our model, with $\left(a_{1},a_{2}\right)=\left(0.862,\,0.138\right)$,
agrees extremely well with the model from \citet{2001MNRAS.328...64N},
with $\left(a_{1},a_{2}\right)=\left(0.85,\,0.15\right)$. This agreement
is quite remarkable since we use data from a different survey and
analyse it with a completely different technique. 

\begin{figure*}
\includegraphics[width=1\textwidth]{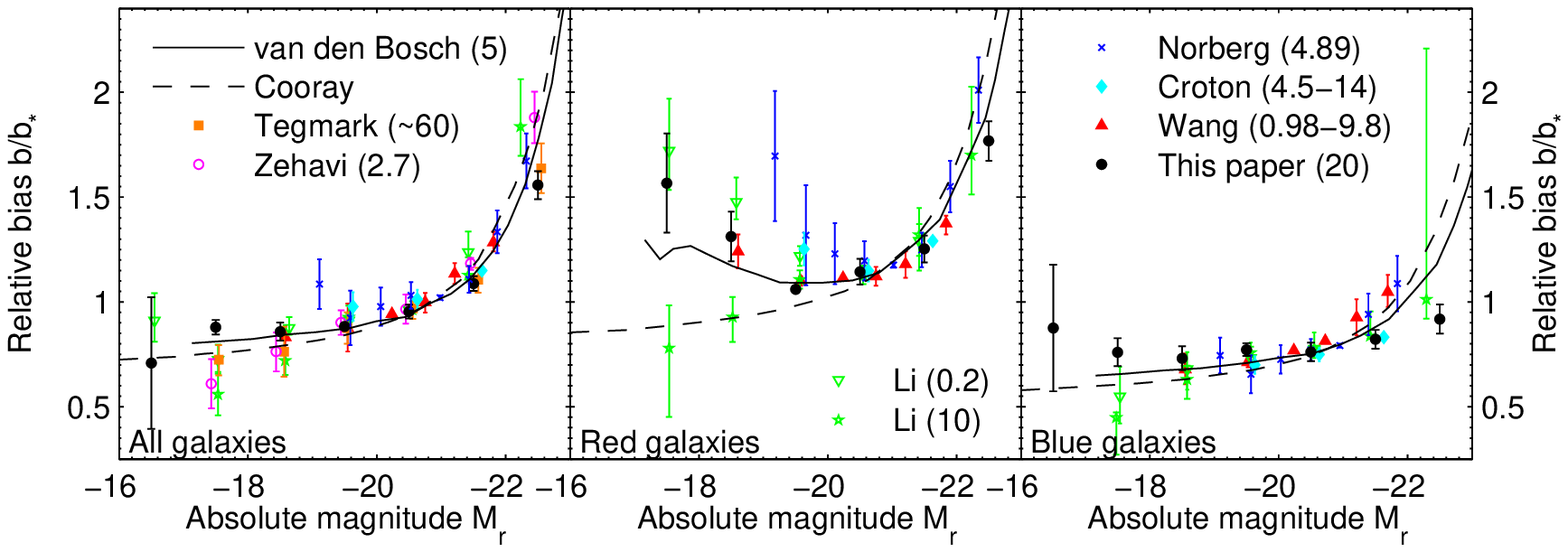}

\caption{\label{fig:comparison_bvmag}Comparison to previous results for the
luminosity dependence of bias for all, red, and blue galaxies. \citet{2002MNRAS.332..827N,2005ApJ...630....1Z,2006MNRAS.368...21L},
and \citet{2007ApJ...664..608W} use correlation function measurements,
\citet{2004ApJ...606..702T} use the power spectrum, and \citet{2007MNRAS.379.1562C}
use counts in cells. To better illustrate the similarities and differences
in the trends as a function of luminosity, we have normalized all
measurements to match our results using the bin closest to $M_{*}=-20.83$.
The error bars shown are all relative: they do not include uncertainties
due to the normalization. Numbers in parentheses denote the scale
in $h^{-1}\rmn{Mpc}$ at which the measurements were done. Also shown
are theoretical models from \citet{2003MNRAS.340..771V} (we show
their model B as a representative example) and \citet{2005MNRAS.363..337C}
-- these are also normalized to match our results at $M_{*}$.}
\end{figure*}

A comparison of our results with previous measurements is shown in
Fig.~\ref{fig:comparison_bvmag} in the left panel. %
 In order to compare our SDSS results with results from 2dFGRS \citep{2002MNRAS.332..827N,2007MNRAS.379.1562C},
we have added a constant factor of $-1.13$ to their quoted values
for $M_{b_{J}}-5\log_{10}h$ in order to line up the value of $M_{*}$
used in \citet{2002MNRAS.332..827N} ($M_{b_{J}}-5\log_{10}h=-19.7$)
with the value used here ($M_{^{01.}r}=-20.83$). Note that this is
necessarily a rough correction since the magnitude in the different
bands varies depending on the spectrum of each galaxy, but this method
provides a reasonable means of comparing the different results. This
plot shows excellent agreement over a wide range of scales, lending
further support to our conclusion that the luminosity-dependent bias
is independent of scale.

We also use equation~\eqref{eq:invert} to calculate $b/b_{*}$ vs.
$M$ for red and blue galaxies separately. To plot the points for
red, blue, and all galaxies on the same $b/b_{*}$ vs. $M$ plot,
we need to determine their relative normalizations. Applying equation~\eqref{eq:invert}
to the red and blue galaxies gives $b_{\alpha,\rmn{red}}/b_{4,\rmn{red}}$
and $b_{\alpha,\rmn{blue}}/b_{4,\rmn{blue}}$, so to normalize the
red-galaxy and blue-galaxy data points to the all-galaxy data points
in Fig.~\ref{fig:lum-bias}, we need to calculate \begin{equation}
\frac{b_{\alpha,\rmn{red}}}{b_{*,\rmn{all}}}=\frac{b_{4,\rmn{all}}}{b_{*,\rmn{all}}}\frac{b_{4,\rmn{red}}}{b_{4,\rmn{all}}}\frac{b_{\alpha,\rmn{red}}}{b_{4,\rmn{red}}}\label{eq:red_norm}\end{equation}
and

\begin{equation}
\frac{b_{\alpha,\rmn{blue}}}{b_{*,\rmn{all}}}=\frac{b_{4,\rmn{all}}}{b_{*,\rmn{all}}}\frac{b_{4,\rmn{red}}}{b_{4,\rmn{all}}}\frac{b_{4,\rmn{blue}}}{b_{4,\rmn{red}}}\frac{b_{\alpha,\rmn{blue}}}{b_{4,\rmn{blue}}}.\label{eq:blue_norm}\end{equation}

The factor $b_{4,\rmn{all}}/b_{*,\rmn{all}}$ is simply the normalization
factor chosen for the above model to give $b\left(M_{*}\right)/b_{*}=1$.
To determine $b_{4,\rmn{red}}/b_{4,\rmn{all}}$, we use best-fitting
values of $\sigma_{1}^{2}$ from the likelihood analysis described
in Section~\ref{sub:Maximum-Likelihood-Method} at the resolution
with cell sizes closest to $20\, h^{-1}\rmn{Mpc}$: $\sigma_{1}$
from the comparison of dimmer and more luminous galaxies in V3 gives
$b_{4,\rmn{all}}$, and similarly $\sigma_{1}$ from the comparison
of blue and red galaxies in L4 gives $b_{4,\rmn{red}}$, so\begin{equation}
\frac{b_{4,\rmn{red}}}{b_{4,\rmn{all}}}=\left(\frac{\sigma_{1,\rmn{red\, vs.\, blue\, L4}}^{2}}{\sigma_{1,\rmn{lum\, vs.\, dim\, V3}}^{2}}\right)^{1/2}.\label{eq:red_v_all}\end{equation}
The blue points are then normalized relative to the red points using
$b_{4,\rmn{blue}}/b_{4,\rmn{red}}$ equal to the measured value of
$b_{\rmn{rel}}$ from the likelihood comparison of blue and red galaxies
in L4. Thus the shapes of the red and blue curves are determined using
the luminosity-dependent bias from the null-buster analysis, but their
normalization uses information from the likelihood analysis as well.

%Thus, the shapes of the red and blue curves are determined using the luminosity-dependent bias, 
%which is independent of scale as seen in Fig.(c) and (d), but their relative normalization 
%is determined by the colour-dependent bias, which has some slight scale dependence as discussed 
%in Section. This means that making this plot at a different scale size would simply induce a 
%small shift in the relative red/blue normalization but would not impact the shapes of the curves.

Splitting the luminosity dependence of the bias by colour reveals
some interesting features. The bias of the blue galaxies shows only
a weak dependence on luminosity, and both luminous ($M\sim-22$) and
dim ($M\sim-17$) red galaxies have slightly higher bias than moderately
bright ($M\sim-20\sim M_{*}$) red galaxies. The previously observed
luminosity dependence of bias, with a weak dependence dimmer than
$L_{*}$ and a strong increase above $L_{*}$, is thus quite sensitive
to the colour selection: the lower luminosity bins contain mostly
blue galaxies and thus show weak luminosity dependence, whereas the
more luminous bins are dominated by red galaxies which drive the observed
trend of more luminous galaxies being more strongly biased. It is
instructive to compare these results with the mean local overdensity
in colour-magnitude space, as in fig.~2 of Blanton et al. (2005a).
Although our bias measurements are necessarily much coarser, it can
be seen that the bias is strongest where the overdensity is largest,
as has been seen previously \citep{2006MNRAS.372.1749A}.

Comparisons of our results to other measurements of luminosity-dependent
bias for red and blue galaxies are shown in Fig.~\ref{fig:comparison_bvmag}
in the middle and right panels. Indications of the differing trends
for red and blue galaxies have been observed in previous work: an
early hint of the upturn in the bias for dim red galaxies was seen
in \citet{2002MNRAS.332..827N}, and recent results \citep{2007ApJ...664..608W}
also indicate higher bias for dim red galaxies at scales $>1\, h^{-1}\rmn{Mpc}$.
However, there is some inconsistency between these results compared
to \citet{2005ApJ...630....1Z} and \citet{2006MNRAS.368...21L} regarding
the dim red galaxies: they find that dim red galaxies exhibit the
strongest clustering on scales $<1\, h^{-1}\rmn{Mpc}$ and luminous
red galaxies exhibit the strongest clustering on larger scales, as
can be seen from the green points in Fig.~\ref{fig:comparison_bvmag}.
This is shown in fig.~14 of \citet{2005ApJ...630....1Z} and fig.~11
of \citet{2006MNRAS.368...21L}. However, we find the dim red galaxies
to have higher bias than $L_{*}$ red galaxies at all the scales we
probe ($2-40\, h^{-1}\rmn{Mpc}$ in this case). This upturn of the
bias for dim red galaxies is present in the halo model-based theoretical
curves from \citet{2003MNRAS.340..771V}, although not in the theoretical
curves from \citet{2005MNRAS.363..337C}. Note also that \citet{2003MNRAS.340..771V}
use the data from \citet{2002MNRAS.332..827N} to constrain their
models so the agreement between the theory and data should be interpreted
with some caution.

Recent measurements of higher order clustering statistics \citep{2007MNRAS.379.1562C}
find the same trends in the clustering strengths of red and blue galaxies,
although they indicate that their linear bias measurement (which should
be comparable to ours) shows the opposite trends -- little luminosity
dependence for red galaxies and a slight monotonic increase for blue
galaxies. However, their luminosity range is much narrower than ours
so the trends are less clear, and placing their data points on Fig.~\ref{fig:comparison_bvmag}
shows that they are in good agreement with our results. %It is also worth noting that these previous studies tend to detect
%at least a somewhat stronger increase in the bias of blue galaxies
%as a function of luminosity than we have measured here.

Previous studies \citep{2002MNRAS.332..827N,2006MNRAS.368...21L,2007ApJ...664..608W}
have also reported a somewhat stronger luminosity dependence of blue
galaxy clustering than we have measured here. As can be seen in Fig.~\ref{fig:comparison_bvmag},
\citet{2002MNRAS.332..827N} and \citet{2007ApJ...664..608W} measure
slightly higher bias for luminous blue galaxies, and \citep{2006MNRAS.368...21L}
measure slightly lower bias for dim blue galaxies. Although the quantitative
disagreement is fairly small, the qualitative trends of the previous
studies imply that the bias of blue galaxies increases with luminosity,
as opposed to our measurement which indicates a lack of luminosity
dependence.

\subsection{\label{sub:Likelihood-Results}Likelihood Results}

To study the luminosity dependence, colour dependence and stochasticity
of bias in more detail, we also apply the maximum likelihood method
described in Section~\ref{sub:Maximum-Likelihood-Method} to all
of the same pairs of samples used in the null-buster test. Due to
constraints on computing power and memory, we perform these calculations
for only three values of the cell size $\elll$ rather than four,
dropping the highest resolution (smallest cell size) shown in Fig.~\ref{fig:angularcells}.
The likelihood analysis makes a few additional assumptions, but provides
a valuable cross-check and also a measurement of the parameter $r_{\rmn{rel}}$
which encodes the stochasticity and non-linearity of the relative
bias.

For each pair of samples, the likelihood function given in equation~\eqref{eq:likelihood}
is maximized with respect to the parameters $\sigma_{1}^{2}$, $b_{\rmn{rel}}$,
and $r_{\rmn{rel}}$ and marginalized over $\sigma_{1}^{2}$ to determine
the best-fitting values of $b_{\rmn{rel}}$ and $r_{\rmn{rel}}$,
with uncertainties defined by the $\Delta\left(2\ln\mathcal{L}\right)=1$
contour in the $b_{\rmn{rel}}$-$r_{\rmn{rel}}$ plane. As we discuss
in Section~ \ref{sub:Comparison-with-null-buster}, the values of
$b_{\rmn{rel}}$ found here are consistent with those determined using
the null-buster test. %{[}without
%the heuristic correction -- but null-buster points could be affected
%in the same way. Could do MC test for null-buster points also], and
%they produce similar results for the luminosity dependence of the
%bias shown in Fig.~\ref{fig:lum-bias}, albeit with slightly larger
%uncertainties. {[}could include plots showing both of these things.]
%
\begin{figure*}
\includegraphics[width=1\textwidth]{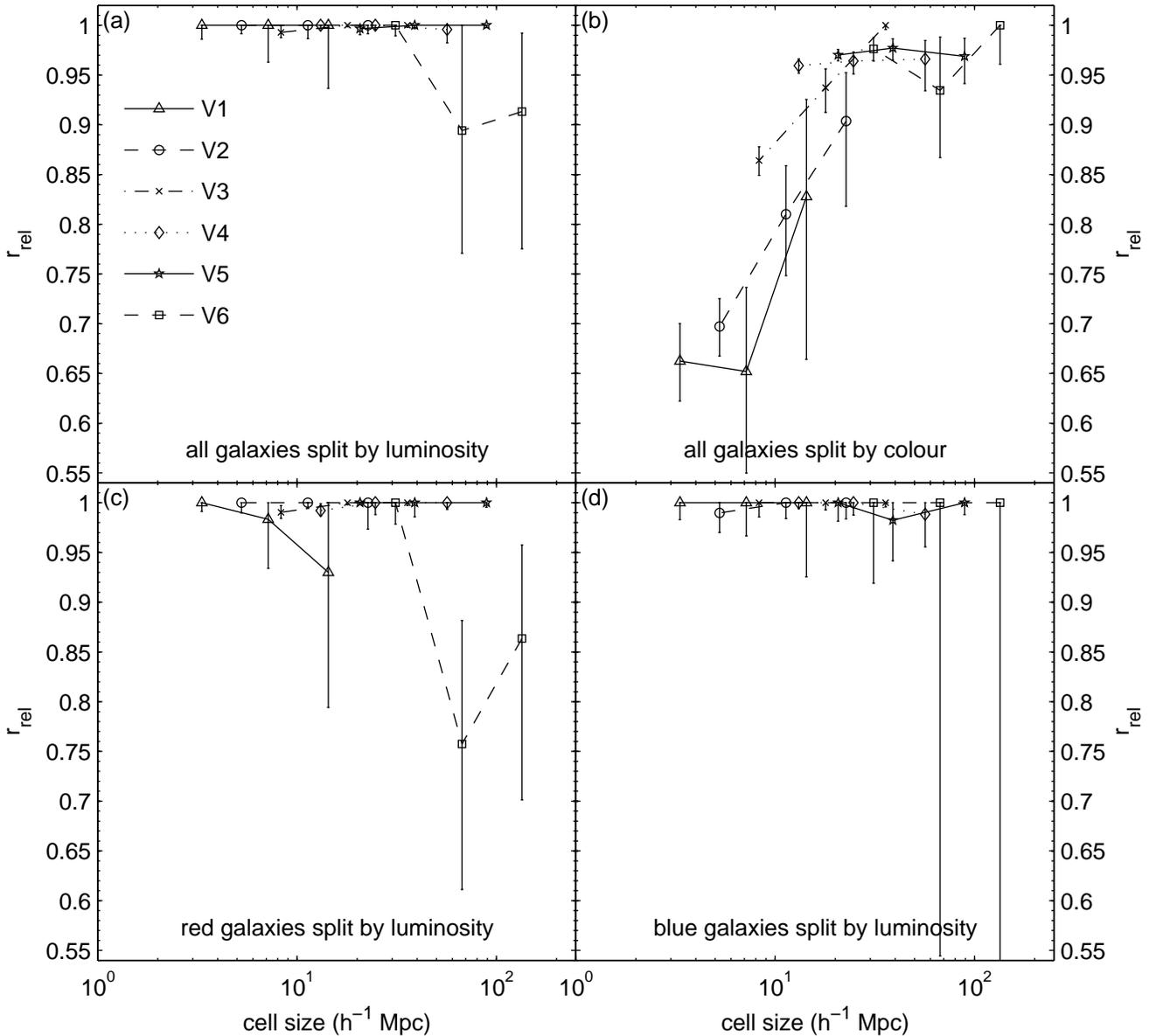}

\caption{\label{fig:rplot}The best-fitting values of the relative cross-correlation
coefficient $r_{\rmn{rel}}$ between pairwise samples. Four different
types of pairwise comparison are illustrated: (a) luminous vs. dim,
(b) red vs. blue, (c) luminous red vs. dim red, and (d) luminous blue
vs. dim blue. The different symbols denote the different comparison
volumes V1-V6. }
\end{figure*}

Figure~\ref{fig:rplot} shows the best-fitting values of $r_{\rmn{rel}}$
as a function of cell size $\elll$. For the comparisons between neighbouring
luminosity bins, the results are consistent with $r_{\rmn{rel}}=1$.
On the other hand, the comparisons between red and blue galaxies give
$r_{\rmn{rel}}<1$, with smaller cell sizes $\elll$ giving smaller
values of $r_{\rmn{rel}}$. This confirms the null-buster result that
the luminosity-dependent bias can be accurately modelled using simple
deterministic linear bias but colour-dependent bias demands a more
complicated model. Also, $r_{\rmn{rel}}$ for the colour-dependent
bias is seen to depend on scale but not strongly on luminosity. In
contrast, $b_{\rmn{rel}}$ (both in the null-buster and likelihood
analyses) depends on luminosity but not on scale.

To summarize, we find that the simple, deterministic model is a good
fit for the luminosity-dependent bias, but the colour-dependent bias
shows evidence for stochasticity and/or non-linearity which increases
in strength towards smaller scales. These results are consistent with
previous detections of stochasticity/non-linearity in spectral-type-dependent
bias \citep{1999ApJ...518L..69T,2000ApJ...544...63B,2005MNRAS.356..456C},
and also agree with \citep{2005MNRAS.356..247W} which measures significant
stochasticity between galaxies of different colour or spectral type,
but not between galaxies of different luminosities. 

\begin{figure}
\includegraphics[width=1\columnwidth]{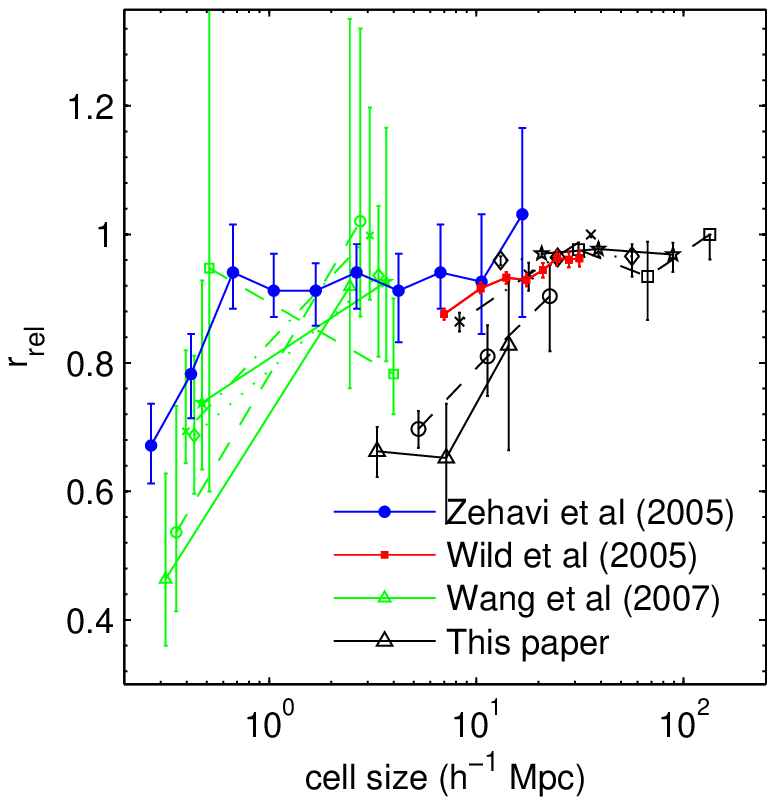}

\caption{\label{fig:comparison_r}Comparison of relative cross-correlation
coefficient $r_{\rmn{rel}}$ between red and blue galaxies as measured
with different techniques. The points from \citet{2005ApJ...630....1Z}
are extracted from cross-correlation measurements between red and
blue galaxies in SDSS with $M_{^{0.1}r}<-21$, and the \citet{2005MNRAS.356..247W}
points are from a counts-in-cells analysis using all 2dFGRS galaxies.
Our results and the \citet{2007ApJ...664..608W} results (also from
SDSS) are separated by luminosity -- symbols are the same as in Fig.~\ref{fig:rplot},
and for the \citet{2007ApJ...664..608W} results, open triangles denote
their dimmest bin ($-19<M_{^{01.}r}<-18$) and open squares denote
their most luminous bin ($-23<M_{^{01.}r}<-21.5$). The length scales
used in \citet{2007ApJ...664..608W} are averages over small scales
($0.16-0.98\, h^{-1}\rmn{Mpc}$) and large scales ($0.98-9.8\, h^{-1}\rmn{Mpc}$)
-- points here are shown in the middle of these ranges and offset
for clarity.}
\end{figure}

We compare our results for $r_{\rmn{rel}}$ for red and blue galaxies
to previous results in Fig.~\ref{fig:comparison_r}. This shows good
agreement between our results and those of \citet{2005MNRAS.356..247W}
($r_{\rmn{lin}}$ from their fig.~11), implying that these results
are quite robust since our analysis uses a different data set, employs
different methods, and makes different assumptions. %

For the results from cross-correlation measurements, however, the
agreement is not as clear. \citet{2005ApJ...630....1Z} find that
the cross-correlation between red and blue galaxies (their fig.~24),
indicates that $r_{\rmn{rel}}$ is consistent with 1 on scales $>1\, h^{-1}\rmn{Mpc}$.
However, it is not clear that this result disagrees with ours, as
their result is for luminous galaxies ($M_{^{0.1}r}<-21$) and and
we do not see a strong indication of $r_{\rmn{rel}}<1$ for our V6
sample ($23<M_{^{0.1}r}<-21$). 

More recent cross-correlation measurements \citep{2007ApJ...664..608W}
do find evidence for stochasticity/non-linearity between red and blue
galaxies at scales $<1\, h^{-1}\rmn{Mpc}$ and also show an indication
that dimmer galaxies have slightly lower values of $r_{\rmn{rel}}$.
Note also that the method of calculating $r_{\rmn{rel}}$ by taking
ratios of cross- and auto-correlation functions as used for \citet{2005ApJ...630....1Z}
and \citet{2007ApJ...664..608W} does not automatically incorporate
the constraint that $\left|r_{\rmn{rel}}\right|\le1$ as our analysis
does, so their error bars are allowed to extend above $r_{\rmn{rel}}=1$
in Fig.~\ref{fig:comparison_r}.

Overall, the counts-in-cells measurements (this paper, \citealt{2005MNRAS.356..247W})
show stronger evidence for stochasticity/non-linearity at larger scales
than the cross-correlation measurements \citep{2005ApJ...630....1Z,2007ApJ...664..608W},
indicating either that there might be some slight systematic variation
between the two methods or that the counts-in-cells method is more
sensitive to these effects.

\section{Conclusions}

\label{Conclusions} To shed further light on how galaxies trace matter,
we have quantified how different types of galaxies trace each other.
We have analysed the relative bias between pairs of volume-limited
galaxy samples of different luminosities and colours using counts-in-cells
at varying length scales. This method is most sensitive to length
scales between those probed by correlation function and power spectrum
methods, and makes point-by-point comparisons of the density fields
rather than using ratios of moments, thereby eliminating sample variance
and obtaining a local rather than global measure of the bias. We applied
a null-buster test on each pair of subsamples to determine if the
relative bias was consistent with % the null hypothesis of 
deterministic linear biasing, and we also performed a maximum-likelihood
analysis to find the best-fitting parameters for a simple stochastic
biasing model.

\subsection{Biasing results}

Our primary results are:

\begin{enumerate}
\item The luminosity-dependent bias for red galaxies is significantly different
from that of blue galaxies: the bias of blue galaxies shows only a
weak dependence on luminosity, whereas both luminous and dim red galaxies
have higher bias than moderately bright ($L_{*}$) red galaxies. 
\item Both of our analysis methods indicate that the simple, deterministic
model is a good fit for the luminosity-dependent bias, but that the
colour-dependent bias is more complicated, showing strong evidence
for stochasticity and/or non-linearity on scales $\la10h^{-1}$ Mpc. 
\item The luminosity-dependent bias is consistent with being scale-independent
over the range of scales probed here $\left(2-160\, h^{-1}\rmn{Mpc}\right)$.
The colour-dependent bias depends on luminosity but not on scale,
while the cross-correlation coefficient $r_{\rmn{rel}}$ depends on
scale but not strongly on luminosity, giving smaller $r_{\rmn{rel}}$
values at smaller scales. 
\end{enumerate}
%In terms of cosmology, the fact that luminosity-dependent bias can
%be modelled with the simple deterministic linear bias model is good
%news -- a luminosity-dependent correction to the power spectrum in
%the manner of \cite{2004ApJ...606..702T} is a reasonable approximation.
%However, the complicated nature of the colour-dependent bias indicates
%that colour selection of a galaxy survey can impact cosmological results,
%especially as future power spectrum measurements push to smaller scales.
%{[}anything else cosmologically relevant to say here?]
% MT: Here's a tweaked version
These results are encouraging from the perspective of using galaxy
clustering to measure cosmological parameters: simple scale-independent
linear biasing appears to be a good approximation on the $\ga60{\, h}^{-1}\rmn{Mpc}$
scales used in many recent cosmological studies (e.g., \citet{2006MNRAS.366..189S,2006PhRvD..74l3507T,2007ApJS..170..377S}).
However, further quantification of small residual effects will be
needed to do full justice to the precision of next-generation data
sets on the horizon. Moreover, our results regarding colour sensitivity
suggest that more detailed bias studies are worthwhile for luminous
red galaxies, which have emerged a powerful cosmological probe because
of their visibility at large distances and near-optimal number density
\citep{2001AJ....122.2267E,2005ApJ...633..560E,2006PhRvD..74l3507T},
since colour cuts are involved in their selection.

\subsection{Implications for galaxy formation}

What can these results tell us about galaxy formation in the context
of the halo model? First of all, as discussed in \citet{2005ApJ...630....1Z},
the large bias of the faint red galaxies can be explained by the fact
that such galaxies tend to be satellites in high mass haloes, which
are more strongly clustered than low mass haloes. Previous studies
have found that central galaxies in low-mass haloes are preferentially
blue, central galaxies in high mass haloes tend to be red, and that
the luminosity of the central galaxy is strongly correlated with the
halo mass \citep{2005MNRAS.358..217Y,2005ApJ...633..791Z}. Our observed
lack of luminosity dependence of the bias for blue galaxies would
then be a reflection of the correlation between luminosity and halo
mass being weaker for blue galaxies than for red ones. %Our observed lack of luminosity
%dependence of the bias for blue galaxies is perhaps an indication
%that the luminosities of blue central galaxies correlate less strongly
%with halo mass than that of red central galaxies. 
% MT: Or that most blue gals aren't central for the mass range where it matters
% (it's only for really massive ones that the bias really shoots up)
Additional work is needed to study this quantitatively and compare
it with theoretical predictions from galaxy formation models.

The detection of stochasticity between red and blue galaxies may imply
that red and blue galaxies tend to live in different haloes -- a study
of galaxy groups in SDSS \citep{2006MNRAS.366....2W} recently presented
evidence supporting this, but this is at odds with the cross-correlation
measurement in \citet{2005ApJ...630....1Z}, which implies that blue
and red galaxies are well-mixed within haloes. The fact that the stochasticity
is strongest at small scales suggests that this effect is due to the
1-halo term, i.e., arising from pairs of galaxies in the same halo,
although some amount of stochasticity persists even for large scales.
However, the halo model implications for stochasticity have not been
well-studied to date. % MT: We intend to pursue this topic in more detail in a future paper.
% The best laid plans of mice and men ... are best not posted to enable subsequent teasing...

%{[}lack of stochasticity in luminosity-dependent bias --> ?]

%{[}lack of scale dependence of b --> requires contribution from both
%1 and 2 halo term?]

%{[}also should compare with semi-analytical theory predictions --
%reading list:\cite{2000MNRAS.318..203S,2001MNRAS.320..289S,2003ApJ...593....1B,2005MNRAS.358..217Y,2005ApJ...633..791Z,2006MNRAS.366....2W,2006MNRAS.371..537W}]

%In summary, 
%our results on galaxy biasing and future work along these
%lines can help place constraints on galaxy formation theories and provide
%a deeper understanding of the systematic uncertainties for future
%cosmological studies using galaxy redshift surveys.

% MT: Less is more:
In summary, our results on galaxy biasing and future work along these
lines should be able to deepen our understanding of both cosmology
(by quantifying systematic uncertainties) and galaxy formation.

\section*{Acknowledgments}

The authors wish to thank Andrew Hamilton for work with the \noun{mangle}
software, and Daniel Eisenstein, David Hogg, Taka Matsubara, Ryan
Scranton, Ramin Skibba, and Simon White for helpful comments. This
work was supported by NASA grants NAG5-11099 and NNG06GC55G, NSF grants
AST-0134999 and 0607597, the Kavli Foundation, and fellowships from
the David and Lucile Packard Foundation and the Research Corporation.

Funding for the SDSS has been provided by the Alfred P.~Sloan Foundation,
the Participating Institutions, the National Science Foundation, the
U.S.~Department of Energy, the National Aeronautics and Space Administration,
the Japanese Monbukagakusho, the Max Planck Society, and the Higher
Education Funding Council for England. The SDSS Web Site is \url{http://www.sdss.org}.

The SDSS is managed by the Astrophysical Research Consortium for the
Participating Institutions. The Participating Institutions are the
American Museum of Natural History, Astrophysical Institute Potsdam,
University of Basel, Cambridge University, Case Western Reserve University,
University of Chicago, Drexel University, Fermilab, the Institute
for Advanced Study, the Japan Participation Group, Johns Hopkins University,
the Joint Institute for Nuclear Astrophysics, the Kavli Institute
for Particle Astrophysics and Cosmology, the Korean Scientist Group,
the Chinese Academy of Sciences (LAMOST), Los Alamos National Laboratory,
the Max-Planck-Institute for Astronomy (MPIA), the Max-Planck-Institute
for Astrophysics (MPA), New Mexico State University, Ohio State University,
University of Pittsburgh, University of Portsmouth, Princeton University,
the United States Naval Observatory, and the University of Washington.

\bibliography{bias,bias1}

\appendix

\section{Consistency Checks}

\subsection{Alternate null-buster analyses}

In order to test the robustness of our results against various systematic
effects, we have repeated the null-buster analysis with four different
modifications: splitting the galaxy samples randomly, offsetting the
pixel positions, using galaxy positions without applying the finger-of-god
compression algorithm, and ignoring the cosmological correlations
between neighbouring cells.

\subsubsection{Randomly split samples}

The null-buster test assumes that the Poissonian shot noise for each
type of galaxy in each pixel is uncorrelated -- i.e., that the matrix
$\mathbfss{N}$ in equation~\eqref{eq:N_and_S} is diagonal -- and
that this shot noise can be approximated as Gaussian. To test the
impact of these assumptions on our results, we repeated the null-buster
analysis using randomly split galaxy samples rather than splitting
by luminosity or colour. For each volume V1-V6, we created two samples
by generating a uniformly distributed random number for each galaxy
and assigning it to sample 1 for numbers $>0.5$ and sample 2 otherwise.

\begin{figure*}
\includegraphics[width=1\textwidth]{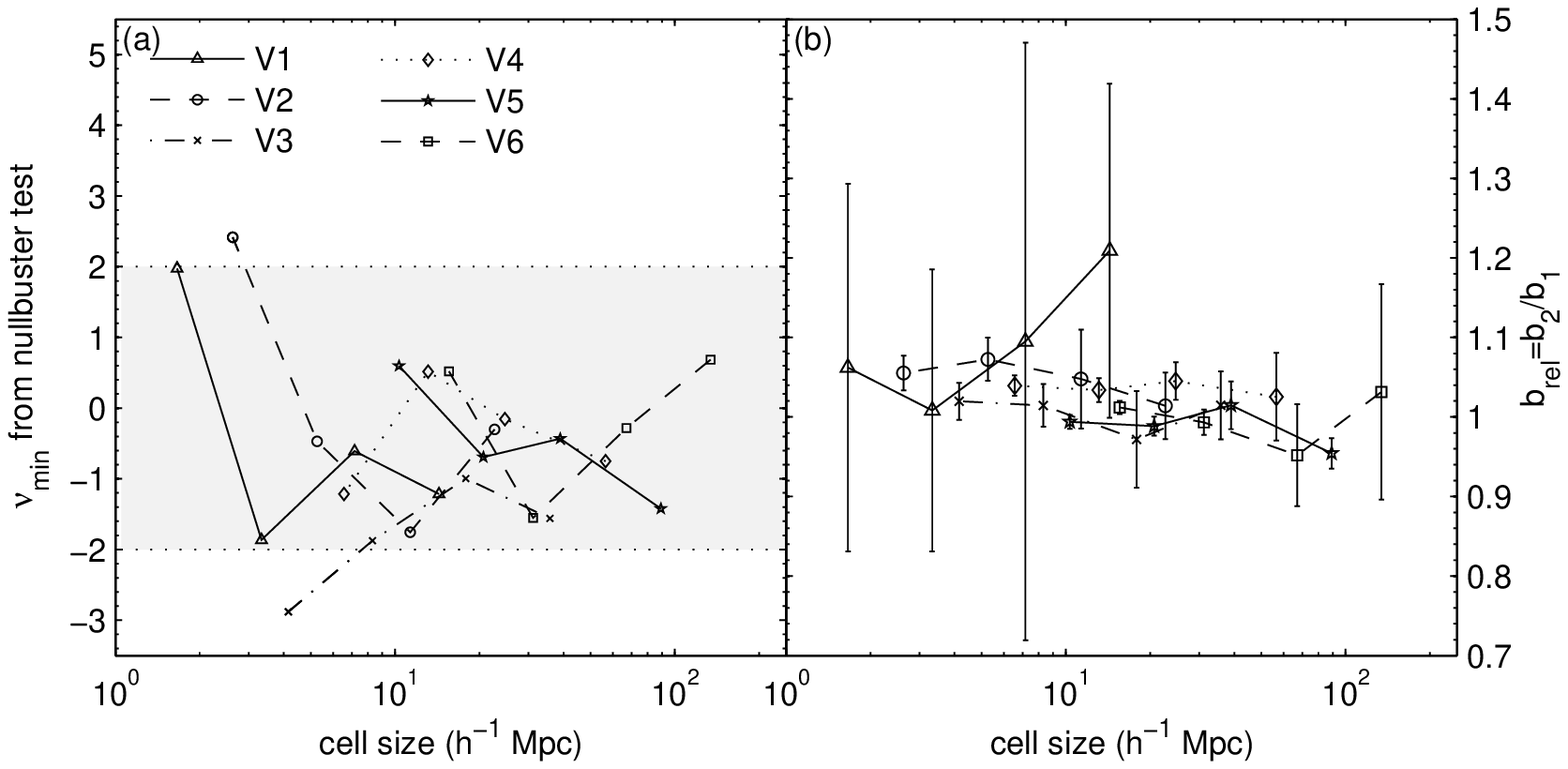}

\caption{\label{fig:randomplot}Null-buster results for randomly split samples.}
\end{figure*}

If the null-buster test is accurate, we expect the pairwise comparison
for the randomly split samples to be consistent with deterministic
linear bias with $b_{\rmn{rel}}=1$. The results are shown in Fig.~\ref{fig:randomplot}
-- we find that, indeed, deterministic linear bias is not ruled out,
with nearly all of the $\nu_{\rmn{min}}$ points falling within $\pm2$.
Furthermore, the measured values of $b_{\rmn{rel}}$ are seen to be
consistent with 1. Thus, we detect no systematic effects due to the
null-buster assumptions.

\subsubsection{Offset pixel positions}

\label{sub:Offset-pixel-positions}

To test if our results are stable against the pixelization chosen,
particularly at large scales where we have a small number of cells,
we shifted the locations on the sky of the angular pixels defining
the cells by half a pixel width in declination. Applying the null-buster
analysis to the offset pixels reveals no significant differences from
the original analysis: the luminosity-dependent comparisons for all,
red, and blue galaxies are still consistent with deterministic linear
bias, and the colour-dependent comparison still shows strong evidence
for stochasticity and/or nonlinearity, especially at smaller scales.

For the all-, red-, and blue-galaxy, luminosity-dependent comparisons,
we also compared the measured values of $b_{\rmn{rel}}$ from the
offset analysis with the original analysis: we took the difference
between the two measured values in each volume at each resolution
and divided this by the larger of the error bars on the two analyses
to determine the number of sigmas by which the two analyses differ.
In order to be conservative, we did not add the error bars from the
two analyses in quadrature, since this would overestimate the error
on the difference if they are correlated and this would make our test
less robust. Note that a fully proper treatment would necessitate
accounting for the correlations between the errors from each analysis,
which we have not done -- the discussions in the this and the following
sections are meant only to serve as a crude reality check.

For the error bars on the original analysis, we used the jackknife
uncertainties described in Section~\ref{sub:Jackknifes}, and for
the offset analysis error bars we use the generalized $\chi^{2}$
uncertainties described in Section~\ref{sub:The-Null-buster-Test}
computed from the offset results. (This is because we did not perform
jackknife resampling for the offset case or the other modified analyses.) 

The results show good agreement: out of a total of 72 measured $b_{\rmn{rel}}$
values, only 4 differ by more than $2\sigma$ (all galaxies in V3
and V5 at the second-smallest cell size, at $2.6\sigma$ and $3.2\sigma$
respectively, and the red galaxies in V3 and V5 at the second-smallest
cell size, at $2.2\sigma$and $3.3\sigma$ respectively). As a rough
test for systematic trends, we also counted the number of measurements
for which the measured value of $b_{\rmn{rel}}$ is larger in each
analysis. We found that in 38 cases the value from the offset analysis
was larger, and in 34 cases the value from the original analysis was
larger, indicating no systematic trends in the deviations.

\subsubsection{No finger-of-god compression}

Our analysis used the finger-of-god compression algorithm from \citet{2004ApJ...606..702T}
with a threshold density of $\delta_{c}=200$. This gives a first-order
correction for redshift space distortions, but it complicates comparisons
to other analyses which work purely in redshift space (e.g. \citealt{2005MNRAS.356..247W})
or use projected correlation functions (e.g. \citealt{2005ApJ...630....1Z}),
particularly at small scales $\left(\lesssim10\, h^{-1}\rmn{Mpc}\right)$
where the effects of virialized galaxy clusters could be significant.
To test the sensitivity of our results to this correction, we repeated
the null-buster analysis with no finger-of-god compression. 

The results show excellent agreement with the original analysis --
the smallest-scale measurements for the colour-dependent comparison
only rule out deterministic linear bias at 30 sigma rather than 40,
but the conclusions remain the same. Additionally, we compared the
measured $b_{\rmn{rel}}$ values to the original analysis as in Section~
\ref{sub:Offset-pixel-positions} and find all 72 measurements to
be within $2\sigma$. In 25 cases, the analysis without finger-of-god
compression gave a larger $b_{\rmn{rel}}$ value, and in 47 cases
the original analysis gave a larger value. This indicates that there
might be a very slight tendency to underestimate $b_{\rmn{rel}}$
if fingers-of-god are not accounted for, but the effect is quite small
and well within our error bars. Thus, the finger-of-god compression
has no substantial impact on our results.

\subsubsection{Uncorrelated signal matrix}

The null-buster test requires a choice of residual signal matrix ${\mathbfss{S}}_{\Delta}$
-- our analysis uses a signal matrix derived from the matter power
spectrum, thus accounting for cosmological correlations between neighbouring
cells. However, these correlations are commonly assumed to be negligible
in other counts-in-cells analyses \citep{2000ApJ...544...63B,2005MNRAS.356..247W,2005MNRAS.356..456C}.
To test the sensitivity to the choice of ${\mathbfss{S}}_{\Delta}$,
we repeated the analysis using ${\mathbfss{S}}_{\Delta}$ equal to
the identity matrix.

Again, we find the results to agree well with the original analysis
and lead to the same conclusions. When comparing the measured $b_{\rmn{rel}}$
values to the original analysis as in Section~ \ref{sub:Offset-pixel-positions},
we find only 4 out of 72 points differing by more than $2\sigma$
(all galaxies in V5 at the smallest cell size, at $-2.6\sigma$, red
galaxies in V4 at the second-smallest and smallest cell size, at $2.2\sigma$
and $2.8\sigma$ respectively, and blue galaxies in V4 at the smallest
cell size, at $-3.1\sigma$). In 39 cases the value of $b_{\rmn{rel}}$
is larger with the uncorrelated signal matrix, and in 33 cases $b_{\rmn{rel}}$
is larger in the original analysis, indicating no strong systematic
effects. Thus we expect our results to be directly comparable to other
counts-in-cells analyses done without accounting for cosmological
correlations.

\section{Uncertainty Calculations}

\subsection{Jackknife uncertainties for null-buster analysis}

\label{sub:Jackknifes}

\begin{figure*}
\includegraphics[width=1\textwidth]{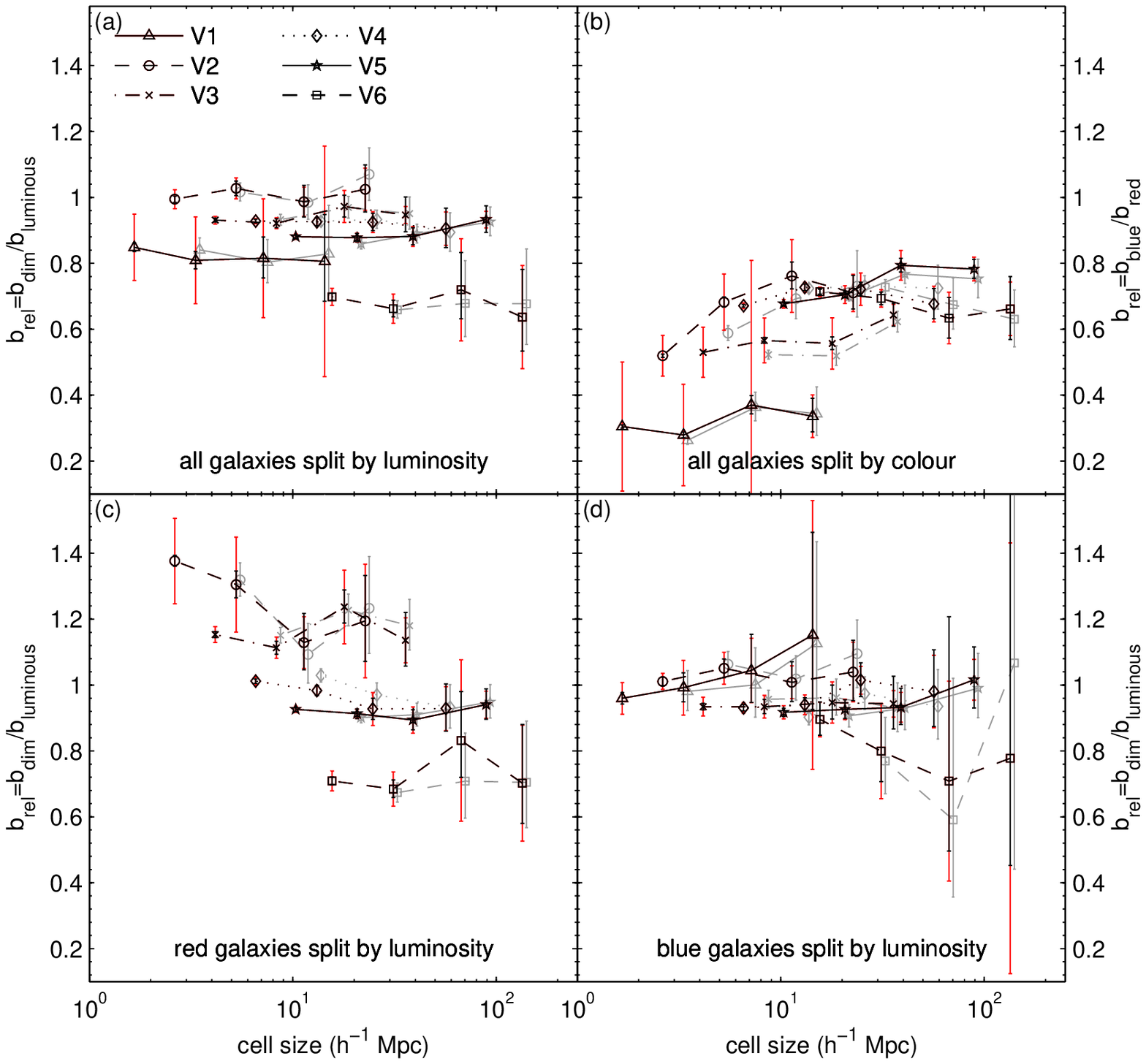}

\caption{\label{fig:likelihoodvnullbuster}Comparison of two methods for calculating
uncertainties on $b_{\rmn{rel}}$ from the null-buster analysis: jackknife
resampling (red) and the generalized $\chi^{2}$ method (black). Also
shown are the results for $b_{\rmn{rel}}$ from the likelihood analysis
(grey).}
\end{figure*}
We use jackknife resampling to calculate the uncertainties for the
null-buster analysis. The concept is as follows: divide area covered
on the sky into $N$ spatially contiguous regions, and then repeat
the analysis $N$ times, omitting each of the $N$ regions in turn.
The covariance matrix for the measured parameters is then estimated
by \begin{equation}
\mathbf{\Sigma}_{\rmn{rel}}^{ij}=\frac{N-1}{N}\sum_{k=1}^{N}\left(b_{\rmn{rel},k}^{i}-\overline{b_{\rmn{rel}}^{i}}\right)\left(b_{\rmn{rel},k}^{j}-\overline{b_{\rmn{rel}}^{j}}\right)\label{eq:jackknife}\end{equation}
where superscripts $i$ and $j$ denote measurements of $b_{\rmn{rel}}$
in different volumes and at different scales, $b_{\rmn{rel},k}^{i}$
denotes the value of $b_{\rmn{rel}}^{i}$ with the $k$th jackknife
region omitted, and $\overline{b_{\rmn{rel}}^{i}}$ is the average
over all $N$ values of $b_{\rmn{rel},k}^{i}$.

For our analysis, we use the 15 pixels at our lowest resolution (upper
left panel in Fig.~\ref{fig:angularcells}) as the jackknife regions.
However, since we use a looser completeness cut at the lowest resolution,
two of these pixels cover an area that is not used at higher resolutions.
Thus we chose not to include these two pixels in our jackknifes since
they do not omit much (or any) area at the higher resolutions. Thus
our jackknife resampling has $N=13$. This technique allows us to
estimate the uncertainties on all of our $b_{\rmn{rel}}$ measurements
as well as the covariance matrix quantifying the correlations between
them. We use these covariances in the model-fitting done in Sections~\ref{sub:scale-dependent}~and~\ref{sub:luminosity-dependent}.

Figure~\ref{fig:likelihoodvnullbuster} shows the uncertainties on
$b_{\rmn{rel}}$ as calculated from jackknife resampling compared
to those calculated with the generalized $\chi^{2}$ method described
in Section~\ref{sub:The-Null-buster-Test}. Overall, the two methods
agree well, but the jackknife method gives larger uncertainties at
the smallest scales and in volume V1. The reason for the large jackknife
uncertainties in volume V1 is because it is significantly smaller
than the other volumes, and it is small enough that omitting a cell
containing just one large cluster can have a substantial effect on
the measured value of $b_{\rmn{rel}}$. Thus the large uncertainties
in V1 reflect the effects of sampling a small volume. Since there
are so few dim red galaxies, these effects are particularly egregious
for the measurements of luminosity-dependent bias of red galaxies
in V1. Thus, based on the jackknife results, we elected to not use
V1 in our analysis of the red galaxies.

\subsection{Likelihood uncertainties}

\label{sub:Likelihood-contour-plots}

\subsubsection{Likelihood contours}

\begin{figure*}
\includegraphics[width=1\textwidth]{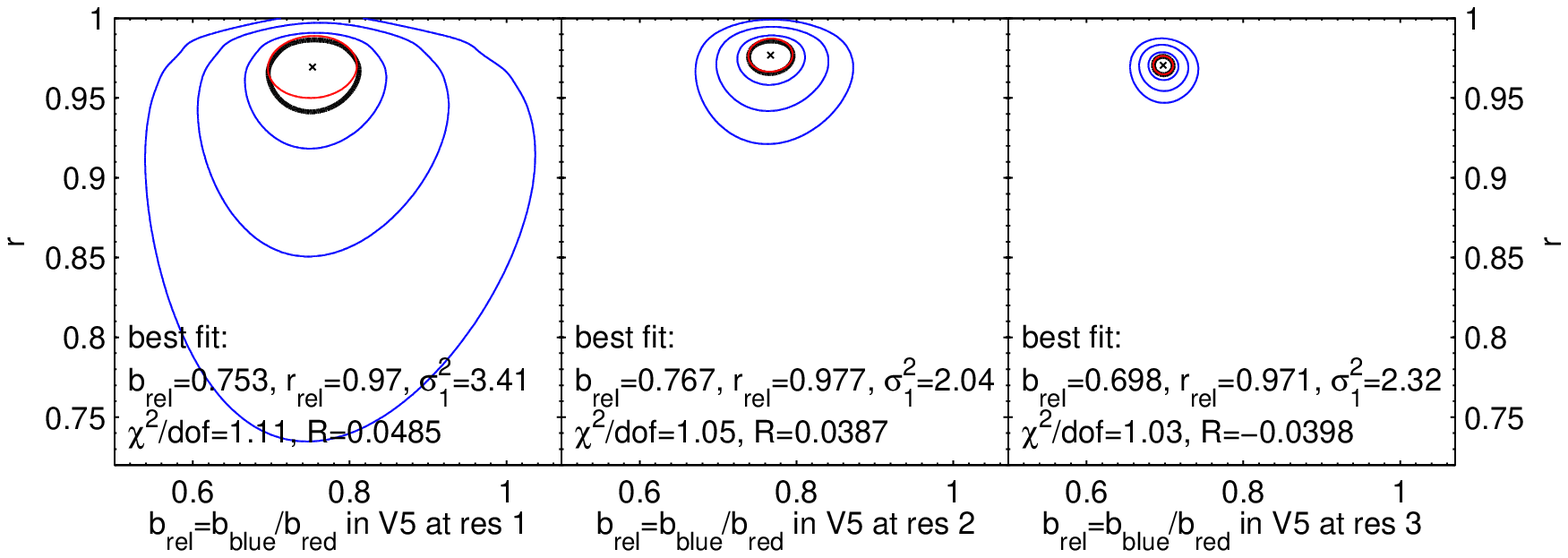}

\caption{\label{fig:likelihoodexamples}Typical contour plots of $\Delta\left(2\ln\mathcal{L}\right)$
for volume V5 for three different resolutions corresponding (from
left- to right-hand-side) to cell sizes of 89, 39, and 21 $h^{-1}\rmn{Mpc}$.
Blue contours denote the $1$, $2$ and $3\sigma$ two-dimensional
confidence regions, and black contours denote the $1\sigma$ one-dimensional
confidence region used for computing error bars on $b_{\rmn{rel}}$
and $r_{\rmn{rel}}$. The red contours denote the error ellipse calculated
from the second-order approximation to $2\ln\mathcal{L}$ at the best-fit
point, marked with a $\times$.}
\end{figure*}
As described in Section~\ref{sub:Maximum-Likelihood-Method}, we
calculate the uncertainty on $b_{\rmn{rel}}$ and $r_{\rmn{rel}}$
for the likelihood method using the $\Delta\left(2\ln\mathcal{L}\right)=1$
contour in the $b_{\rmn{rel}}$-$r_{\rmn{rel}}$ plane after marginalizing
over $\sigma_{1}^{2}$. This means that for each comparison volume
and at each resolution, we calculate $\mathcal{L}$ from equation~\eqref{eq:likelihood}
over a grid of $b_{\rmn{rel}}$ and $r_{\rmn{rel}}$ values and maximize
$2\ln\mathcal{L}$ with respect to $\sigma_{1}^{2}$ at each grid
point. This gives us a 2-dimensional likelihood function, which we
then maximize to find the best-fit values for $b_{\rmn{rel}}$ and
$r_{\rmn{rel}}$. Uncertainties are calculated using the function
\begin{equation}
\Delta\left(2\ln\mathcal{L}\left(b_{\rmn{rel},}r_{\rmn{rel}}\right)\right)\equiv2\ln\mathcal{L}\left(b_{\rmn{rel},}^{\rmn{max}}r_{\rmn{rel}}^{\rmn{max}}\right)-2\ln\mathcal{L}\left(b_{\rmn{rel},}r_{\rmn{rel}}\right).\label{eq:contour}\end{equation}
Typical contour plots of this function for volume V5 at each of the
three cell sizes used are shown in Fig.~\ref{fig:likelihoodexamples}.

We define 1- and 2-dimensional confidence regions using the standard
procedures detailed in \citet{1992nrca.book.....P}, using $\Delta\left(2\ln\mathcal{L}\right)$
as an equivalent to $\Delta\chi^{2}$: the $1\sigma$ (68.3\%) 1-dimensional
confidence region is given by $\Delta\left(2\ln\mathcal{L}\right)=1$,
so we define our error bars on $b_{\rmn{rel}}$ and $r_{\rmn{rel}}$
by projecting the $\Delta\left(2\ln\mathcal{L}\right)=1$ contour
(shown in black in Fig.~\ref{fig:likelihoodexamples}) onto the $b_{\rmn{rel}}$
and $r_{\rmn{rel}}$ axes. For illustrative purposes we also show
the $1\sigma$, $2\sigma$, and $3\sigma$ (68.3\%, 95.4\%, and
99.73\%) 2-dimensional confidence regions in these plots, given by
$\Delta\left(2\ln\mathcal{L}\right)=2.30$, $6.17$, and $11.8$ respectively.

To check the goodness of fit, we also compute an effective value of
$\chi^{2}$: \begin{equation}
\chi_{\rmn{eff}}^{2}\equiv-2\ln\mathcal{L}-\ln\left|\mathbfss{C}\right|-2n\ln\left(2\pi\right),\label{eq:chi_eff}\end{equation}
where $\mathbfss{C}$ is given by equation~\eqref{eq:cov_matrix}
and $n$ is the number of cells. If our model is a good fit, the value
of $\chi_{\rmn{eff}}^{2}$ at the best fit parameter values should
be close to the number of degrees of freedom, given by $\rmn{dof}=2n-2$
($2n$ data points for type 1 and 2 galaxies in each cell minus 2
parameters $b_{\rmn{rel}}$ and $r_{\rmn{rel}}$). We calculated $\chi_{\rmn{eff}}^{2}/\rmn{dof}$
for each volume and resolution and found they all lie quite close
to 1, ranging from a minimum value of $0.678$ to a maximum value
of $1.11$. Thus this test indicates our model is a good fit.

The uncertainties on $b_{\rmn{rel}}$ and $r_{\rmn{rel}}$ could perhaps
be calculated more accurately using jackknife resampling as we did
for the null-buster case; however, repeating the analysis for each
jackknife sample is computationally prohibitive since performing all
the calculations for just one likelihood analysis took several months
of CPU time.

\subsubsection{$b_{\rmn{rel}}$-$r_{\rmn{rel}}$ covariance matrices}

Alternatively, we can calculate the uncertainties using the parameter
covariance matrix at the best fit parameter values, as is commonly
done in $\chi^{2}$ analyses. The Hessian matrix of second derivatives
is given by \begin{equation}
\mathbfss{H}\equiv\left(\begin{array}{cc}
\frac{d^{2}\left(2\ln\mathcal{L}\right)}{db_{\rmn{rel}}^{2}} & \frac{d^{2}\left(2\ln\mathcal{L}\right)}{db_{\rmn{rel}}dr_{\rmn{rel}}}\\
\frac{d^{2}\left(2\ln\mathcal{L}\right)}{db_{\rmn{rel}}dr_{\rmn{rel}}} & \frac{d^{2}\left(2\ln\mathcal{L}\right)}{dr_{\rmn{rel}}^{2}}\end{array}\right)\label{eq:hessian}\end{equation}
and the parameter covariance matrix is given by \begin{equation}
{\mathbfss{C}}_{\rmn{param}}\equiv2{\mathbfss{H}}^{-1}=\left(\begin{array}{cc}
\sigma_{b_{\rmn{rel}}}^{2} & \sigma_{{b_{\rmn{rel}}r}_{\rmn{rel}}}^{2}\\
\sigma_{{b_{\rmn{rel}}r}_{\rmn{rel}}}^{2} & \sigma_{r_{\rmn{rel}}}^{2}\end{array}\right).\label{eq:param_cov}\end{equation}
Thus the uncertainties are given by $\sigma_{b_{\rmn{rel}}}^{2}$
and $\sigma_{r_{\rmn{rel}}}^{2}$ with this method. This is equivalent
to approximating the likelihood function $2\ln\mathcal{L}$ with its
second-order Taylor series about the best-fit point, and it defines
an error ellipse that approximates the $\Delta\left(2\ln\mathcal{L}\right)=1$
contour. These error ellipses are shown in Fig.~\ref{fig:likelihoodexamples}
in red, and are seen to be in close agreement with the true $\Delta\left(2\ln\mathcal{L}\right)=1$
contours.

This method also allows us to measure the correlation between $b_{\rmn{rel}}$
and $r_{\rmn{rel}}$ by calculating the correlation coefficient, given
by \begin{equation}
R\equiv\frac{\sigma_{{b_{\rmn{rel}}r}_{\rmn{rel}}}^{2}}{\left(\sigma_{b_{\rmn{rel}}}^{2}\sigma_{r_{\rmn{rel}}}^{2}\right)^{1/2}}.\label{eq:corr_coeff}\end{equation}
$R$ will fall between -1 (perfectly anti-correlated) and 1 (perfectly
correlated). Effectively this measures the tilt of the error ellipse
in the $b_{\rmn{rel}}$-$r_{\rmn{rel}}$ plane. Overall we find the
values of $R$ to be quite small -- typically $\left|R\right|\sim0.05$
-- indicating no large correlations between $b_{\rmn{rel}}$ and $r_{\rmn{rel}}$.
Out of the 72 points we calculate, only 6 have $\left|R\right|>0.3$.
The cases with the largest $R$ values are for blue galaxies in volume
V6, where the uncertainties are quite large due to the small number
of bright blue galaxies and the error ellipses are not good approximations
to the likelihood contours anyway -- thus these few cases with large
$R$ are not overly concerning. 

%; these 6 points are detailed in Table~ \ref{tab:corrcoeffs}%\begin{table}
%\caption{\label{tab:corrcoeffs}Cases which have a cross-correlation coefficient $\left|R\right|>0.3$}
%\begin{tabular}{lccc} \hline  & & & \tabularnewline Comparison& Volume& Cell size & $R$\tabularnewline \hline  All 
%galaxies split by luminosity& V6& $67\, h^{-1}\rmn{Mpc}$& 0.359\tabularnewline All galaxies split by luminosity& V6& 
%$31\, h^{-1}\rmn{Mpc}$& 0.369\tabularnewline Red galaxies split by luminosity& V6& $31\, h^{-1}\rmn{Mpc}$& 
%0.396\tabularnewline Blue galaxies split by luminosity& V5& $21\, h^{-1}\rmn{Mpc}$& 0.386\tabularnewline Blue 
%galaxies split by luminosity& V6& $67\, h^{-1}\rmn{Mpc}$& 0.812\tabularnewline Blue galaxies split by luminosity& 
%V6& $31\, h^{-1}\rmn{Mpc}$& 0.640\tabularnewline \hline \end{tabular} \end{table} 

\subsubsection{Comparison with null-buster results}

\label{sub:Comparison-with-null-buster}

Finally, we compare the results for $b_{\rmn{rel}}$ from the likelihood
method to the results from the null-buster analysis in Fig.~\ref{fig:likelihoodvnullbuster},
with the likelihood points shown in grey. As can be seen in this plot,
the likelihood and null-buster values for $b_{\rmn{rel}}$ agree within
the uncertainties, even for the colour-dependent bias where the null-buster
values are not necessarily accurate since deterministic linear bias
is ruled out. Thus our two analysis methods are in excellent agreement
with each other.
\end{document}